\documentclass[fleqn,usenatbib]{mnras}

\usepackage{newtxtext,newtxmath}

\usepackage[T1]{fontenc}

\DeclareRobustCommand{\VAN}[3]{#2}
\let\VANthebibliography\thebibliography
\def\thebibliography{\DeclareRobustCommand{\VAN}[3]{##3}\VANthebibliography}

\usepackage{graphicx}	
\usepackage{amsmath}	

\usepackage{booktabs}  
\usepackage[authoryear]{natbib}
\usepackage{chngcntr}  
\usepackage{hyperref}
\hypersetup{
    citecolor=blue,
    colorlinks=true,
    linkcolor=blue,
    filecolor=magenta,      
    urlcolor=blue,
    pdfpagemode=FullScreen,
    }
\usepackage{CJKutf8}
\usepackage{threeparttable}
\usepackage{mathtools} 
\usepackage{multirow}

\newcommand{\sect}[1]{Section~\ref{#1}}
\newcommand{\fig}[1]{Fig.~\ref{#1}}
\newcommand{\tab}[1]{Table~\ref{#1}}
\newcommand{\equ}[1]{Eq.~(\ref{#1})}
\newcommand{\apdx}[1]{Appendix~\ref{#1}}

\defcitealias{rs75}{RS}
\defcitealias{hww+25}{H25}

\title[Co-existing periodicities in PSR J1901$-$0906]{The Thousand-Pulsar-Array programme on MeerKAT -- XX. Co-existing interacting periodicities and polarization modulation in PSR J1901$-$0906}

\author[J.~A. Hsu et al.]
{Jui-An Hsu (\begin{CJK*}{UTF8}{bsmi}許睿安\end{CJK*})$^{1}$\thanks{E-mail: jui-an.hsu@manchester.ac.uk},
Patrick Weltevrede$^{1}$,
Geoff Wright$^{1}$,
Michael J. Keith$^{1}$,
Aris Karastergiou$^{2}$,
\newauthor and Xiaoxi Song$^{3}$
\\
$^{1}$Jodrell Bank Centre for Astrophysics, Department of Physics and Astronomy, University of Manchester, Manchester M13 9PL, UK \\
$^{2}$Department of Astrophysics, University of Oxford, Denys Wilkinson Building, Keble Road, Oxford OX1 3RH, UK\\
$^{3}$ASTRON, The Netherlands Institute for Radio Astronomy, Oude Hoogeveensedijk 4, 7991 PD, Dwingeloo, the Netherlands \\
}

\date{Accepted XXX. Received YYY; in original form ZZZ}

\pubyear{\the\year{}}

\begin{document}
\label{firstpage}
\pagerange{\pageref{firstpage}--\pageref{lastpage}}

\maketitle

\begin{abstract}
We present a comprehensive analysis of the complex subpulse modulation in PSR~J1901$-$0906 using data from the Thousand-Pulsar-Array (TPA) programme. 
Our analysis reveals the coexistence of three distinct periodicities in the subpulse modulation of both profile components, which obey a beat relation. 
We also detect additional spectral beat features caused by the modulation of these co-existing periodicities by investigating the Fourier phase correlation in two-dimensional fluctuation spectra.
Moreover, the single-pulse polarization states are shown to be modulated and synchronized with the three co-existing periodicities observed in total intensity. 
Such synchronization has been reported for only a small number of pulsars, but this discovery marks the only one where this synchronization involves multiple co-existing periodicities.
The modulation beats are similar to, yet more complex than, the beat system previously identified in PSR~J1514$-$4834 with higher-order harmonics playing a significant role.
The coexistence of the observed beat-like periodicities in PSR~J1901$-$0906 implies that they are not independent, but rather originate from an underlying, interconnected physical system within the pulsar magnetosphere.
Building upon the same geometric framework previously applied to PSR~J1514$-$4834, we demonstrate that a feedback system with interacting rotating carousels successfully reproduces the complex subpulse modulations observed in PSR~J1901$-$0906 and naturally explains the beat relations.
\end{abstract}




\section{Introduction}
\label{sec:intro}

The phenomenon of drifting subpulses, manifesting itself as a systematic shift of emission in pulse longitude from one rotation to the next \citep{dc68}, is common in the pulsar population \citep[e.g.][]{ran86, wes06, wse07, bmm+19, sws+23} and remains one of the most insightful probes into the electrodynamics of the pulsar magnetosphere.
This phenomenon is often interpreted within the framework of the rotating carousel model \citep[][hereafter the \citetalias{rs75} model]{rs75}, which attributes the emission to localized plasma discharges (known as sparks) that circulate around the magnetic axis due to an $\boldsymbol{E} \boldsymbol{\times} \boldsymbol{B}$ drift.

While this model successfully explains the regular drifting subpulses observed in many pulsars, accounting for various observed complexities of emission variability is problematic.
Recent studies have revealed subpulse modulation far more complex than a single rotating carousel can accommodate, such as bi-drifting \citep[e.g.][]{clm+05, PSRSALSA, wkw+25}, main pulse and interpulse interaction \citep[e.g.][]{wws07, swv+26}, and multiple co-existing periodicities \citep[][hereafter \citetalias{hww+25}]{hww+25}.
In the latter work, the slow drifting subpulses in PSR~J1514$-$4834 are shown to be amplitude-modulated by a rapid periodicity, resulting in two additional beat frequencies. This challenges the standard \citetalias{rs75} model which typically predicts a single modulation period associated with the circulation of the emission pattern.
These two co-existing periodicities interact with each other in the form of a ‘beat system’, producing two extra detectable beat frequencies in the fluctuation spectra.
This suggests that these spectral modulation features may actually stem from a single, interconnected physical mechanism.

Here we identify a more complex example of a beat system in PSR~J1901$-$0906, which has similarities to that of PSR~J1514$-$4834.
Discovered by the Parkes southern pulsar survey \citep{lml+98}, this pulsar is characterized by a rotation period of $P\approx1.78$~s and a period derivative of $\dot{P}\approx1.6\times10^{-15}~\rm s~s^{-1}$ \citep{kjk+24}.
In the $P$--$\dot{P}$ diagram, it is located near the pulsar death line \citep[e.g.][]{zhm00}, with a low spin-down energy loss rate of $\dot{E} \approx 1.1\times10^{31}~\rm erg~s^{-1}$ and a characteristic age of $\tau_c \approx 1.7\times10^{7}$~yr.
Previous studies have characterized the modulation of this pulsar as highly variable \citep{wes06, wse07, bmm+19, sws+23}, with \citet{njm+17} classifying it as a drift mode changer that switches between three distinct drift rates.

However, our analysis of data from the MeerKAT Thousand-Pulsar-Array (TPA) programme \citep{jkk+20} suggests that this complex modulation should not be interpreted as transitions between drift modes of which only one exists at a given time. Instead, three periodicities co-exist and obey a beat relation. Such beat frequencies in pulsars are unusual, but not unique. For example, the beat frequencies in PSR~B0943$+$10 are explained by a rotating carousel with subbeams of fixed but unequal intensities \citep{dr01}, while the beat frequencies in PSR~J1514$-$4834 require an alternative model such as two interacting carousels \citepalias{hww+25}.
In a broader context, beat frequencies have been widely invoked across diverse astrophysical systems, wavebands, and time-scales.
Examples include the quasi-periodic oscillations (QPOs) observed in neutron star X-ray binaries on millisecond to second time-scales \citep[e.g.,][]{as85, vjv+85, szs+96, mlp98}, as well as optical flux modulations in white dwarf (WD) binaries occurring on time-scales of minutes to hours \citep[e.g.,][]{war86, reh+25}.
Similar beat frequencies span the ultraviolet to radio regimes in WD pulsars \citep{mgh+16, cpg+25} and extend over much longer time-scales on the order of hours in long-period radio transients \citep{hrh+26}.
Furthermore, precession can also give rise to beat frequencies, as observed in WD binaries \citep{whi88, tzn07, ntz+13} and central compact objects \citep[CCOs,][]{mue26}.

In this work, we demonstrate that the three co-existing periodicities and their beat relation observed in PSR~J1901$-$0906 can be explained by the feedback model of \cite{wri22}, which has been previously applied to PSR~J1514$-$4834.
This model requires two carousels interacting with each other by exchanging particle flows. 
The feedback system may be intrinsic to the pulsar magnetosphere itself or may result from possible external influences such as fallback debris disks \citep{md81, cs08}, the interstellar medium \citep{wri79a, che85}, as well as the precipitation of trapped plasma from radiation belts near the light cylinder \citep{lm07}.
In such a feedback system, the two interacting carousels could be situated at very different locations in the magnetosphere. For example, in addition to the normal radio emission from the polar caps as described in the \citetalias{rs75} model, 
it has been proposed that radio emission may also originate in the outer magnetosphere near the light cylinder \citep[e.g.][]{dhr04, bpd+15}, or even from the current sheet beyond the light cylinder \citep[e.g.][]{kj26}.

Complementary to the study of drifting subpulses, the analysis of polarization offers constraints on the emission geometry and propagation effects in the magnetosphere.
The linear polarization position angle (PA) in individual pulses often exhibits a bimodal distribution separated by $90^\circ$, corresponding to two orthogonal polarization modes (OPMs) \citep[e.g.,][]{em69, mth75, brc76}.
They are usually interpreted as the two natural propagation modes arising from birefringence in the strongly magnetized plasma: the extraordinary (X) mode and the ordinary (O) mode \citep{cp76, ms77, ab86}.
In this work, we find that the dominance of the two OPMs switches periodically and synchronously with the drifting subpulses for PSR~J1901$-$0906.
Such synchronization has only been identified in a small number of pulsars: B0809$+$74 \citep{thh+71, rrs+02, edw04}, B1237$+$25 \citep{rr03}, B0320$+$39 \citep{edw04}, B0818$-$13 \citep{edw04}, B0031$-$07 \citep{iwj+20}, and B1919$+$21 \citep{ptv+22, cas23}.
This suggests that the carousel structures are subjected to propagation effects. 

The structure of this paper is as follows.
In \sect{sec:obs}, we describe the MeerKAT observations and the data processing pipeline.
\sect{sec:J1901} details the analysis of the subpulse modulation, establishing the coexistence of the three periodicities and demonstrating that a simple beat system similar to that of PSR~J1514$-$4834 is insufficient to explain the complexity observed in PSR~J1901$-$0906.
In \sect{sec:pol}, we extend this analysis to the full polarization state, revealing the synchronization between the total intensity modulation and the polarization properties.
The nulling behaviour of PSR~J1901$-$0906 is examined in \sect{sec:null}.
Finally, we discuss the implications of these findings in \sect{sec:discussion}, followed by our conclusions in \sect{sec:conclusion}.

\section{Observation and data processing}
\label{sec:obs}

As part of the TPA programme, PSR~J1901$-$0906 has been monitored with the MeerKAT telescope at $L$-band (856 to 1712 MHz).
The data analysed in this paper were drawn from the TPA ‘census data set’ with the full available array (in this case 59 of 64 antennas participated) and from regular monthly observations conducted with a $\sim$32-dish sub-array \citep{swk+21}.
In total, 7076 pulsar rotations from 65 observations at $L$-band between October 2019 and April 2026 are used in this analysis, as summarized in \apdx{apdx:obs_NF}.

Full-polarization data were recorded in ‘search mode’ with the Pulsar Timing User Supplied Equipment (PTUSE) backend \citep{bja+20} and processed off-line using the MeerTime Single Pulse pipeline \citep[MTSP,][]{kwo+26}.
This pipeline applies polarization calibration and removes frequency channels contaminated by radio frequency interference (typically about 20 to 30 per cent for this pulsar).
Then, the data were folded at the pulsar period using {\sc DSPSR} \citep{DSPSR} into 1024 pulse longitude bins for each single pulse. The frequency-averaged pulse stacks were generated following \cite{sws+23}.
Further details of the observing setup and data processing can be found in \citetalias{hww+25}.

\section{The subpulse modulation of PSR~J1901$\,$--$\,$0906}
\label{sec:J1901}

\subsection{Three co-existing periodicities}
\label{sec:F123}

\begin{figure}
\centering
\includegraphics[width=0.8\hsize]{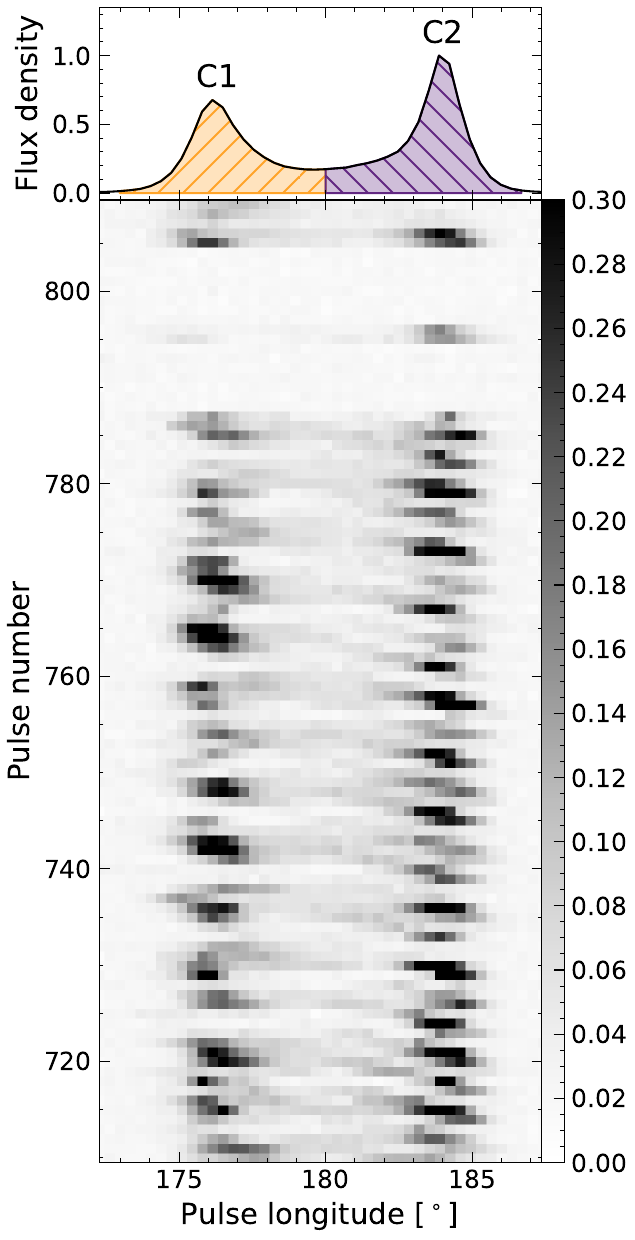}
  \caption{The normalized pulse profile (top) and a section of a pulse stack (bottom) for PSR~J1901$-$0906. The average pulse profile is obtained from all available TPA $L$-band data, while the pulse stack is from the census data set. The leading profile component (C1) and the trailing profile component (C2) are highlighted as shaded regions in different colours in the top panel. The flux density in the pulse stack is clipped at 30 per cent of the maximum flux density for all samples, as indicated by the colour bar. Nulls are detected between pulse numbers 788--794 and 797--804.}
  \label{fig:pulsestack}
\end{figure}

\fig{fig:pulsestack} shows the average pulse profile and a section of the pulse stack for PSR~J1901$-$0906.
Two emission components, C1 and C2, are separated by about $8^\circ$ in the double-peaked pulse profile.
Both components display subpulses with clear negative drift characterized by diagonal emission bands moving toward earlier pulse longitudes.
The emission in C2 is slightly stronger with steadier drifting subpulses, whereas that in C1 is less regular.

\begin{figure}
    \centering
    \includegraphics[width=\linewidth]{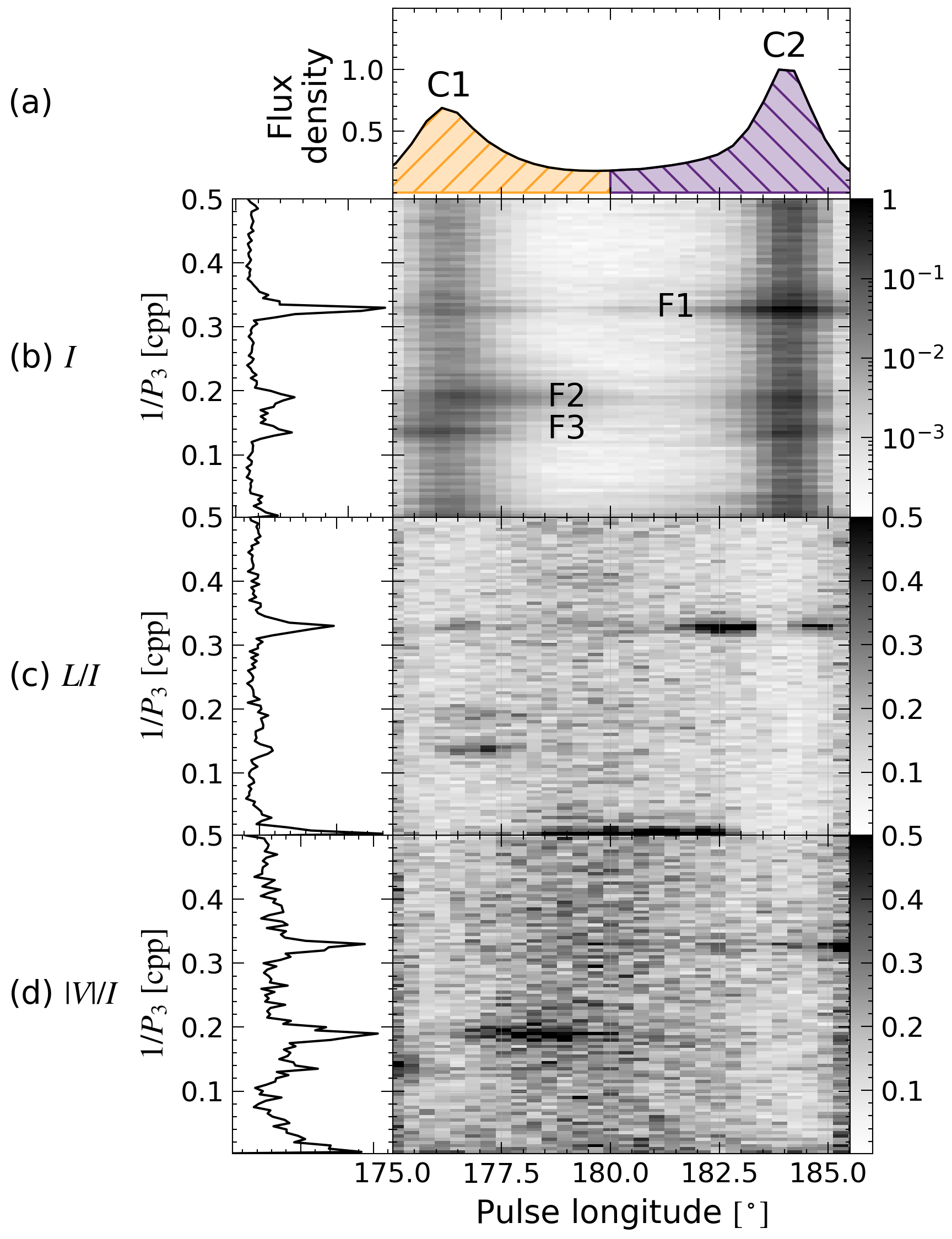}
    \caption{(a) The normalized pulse profile and the power LRFS based on: (b) Stokes $I$, (c) fractional linear polarization intensity $L/I$, and (d) fractional circular polarization intensity $|V|/I$. For the latter two quantities, the sample values within the pulse stacks are restricted to the interval $[0, 1]$ to avoid noise dominating.
    Only data from October 2019 to March 2021 (3600 pulsar rotations in total, including nulls) are used. 
    The FFT length is 200 pulse periods. 
    The DC row ($1/P_3 = 0$) has been set to zero.
    All spectra are normalized separately such that the maximum power equals one. 
    In panel (b) a logarithmic grey scale is used, while in panels (c) and (d) it is linear and capped at 50 per cent of the highest magnitude bin to highlight weaker features.
    Three spectral features, F1, F2, and F3, are labelled in panel (b).
    }
    \label{fig:LRFS}
\end{figure}

\begin{figure}
\centering
\includegraphics[width=0.9\hsize]{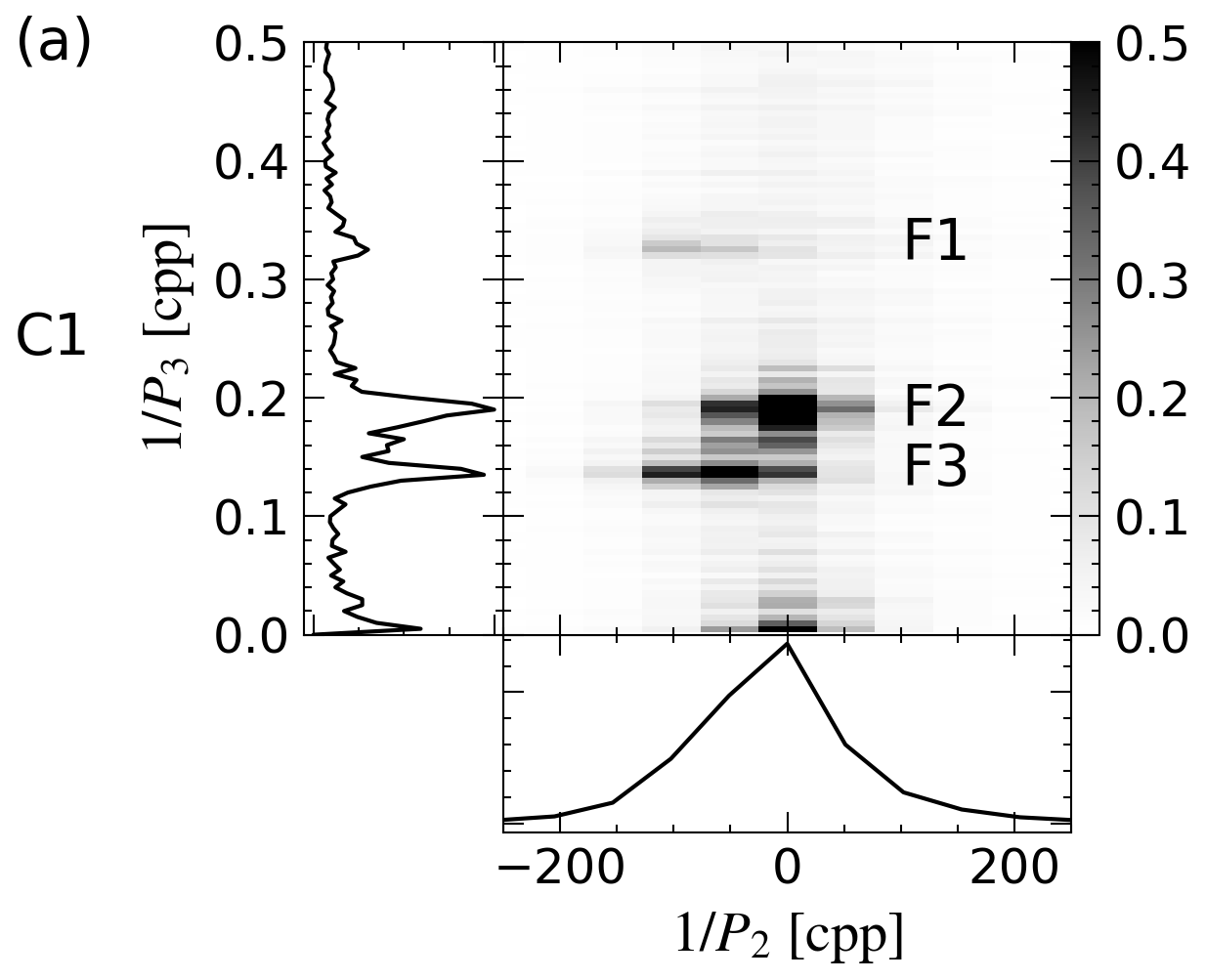} \\
\includegraphics[width=0.9\hsize]{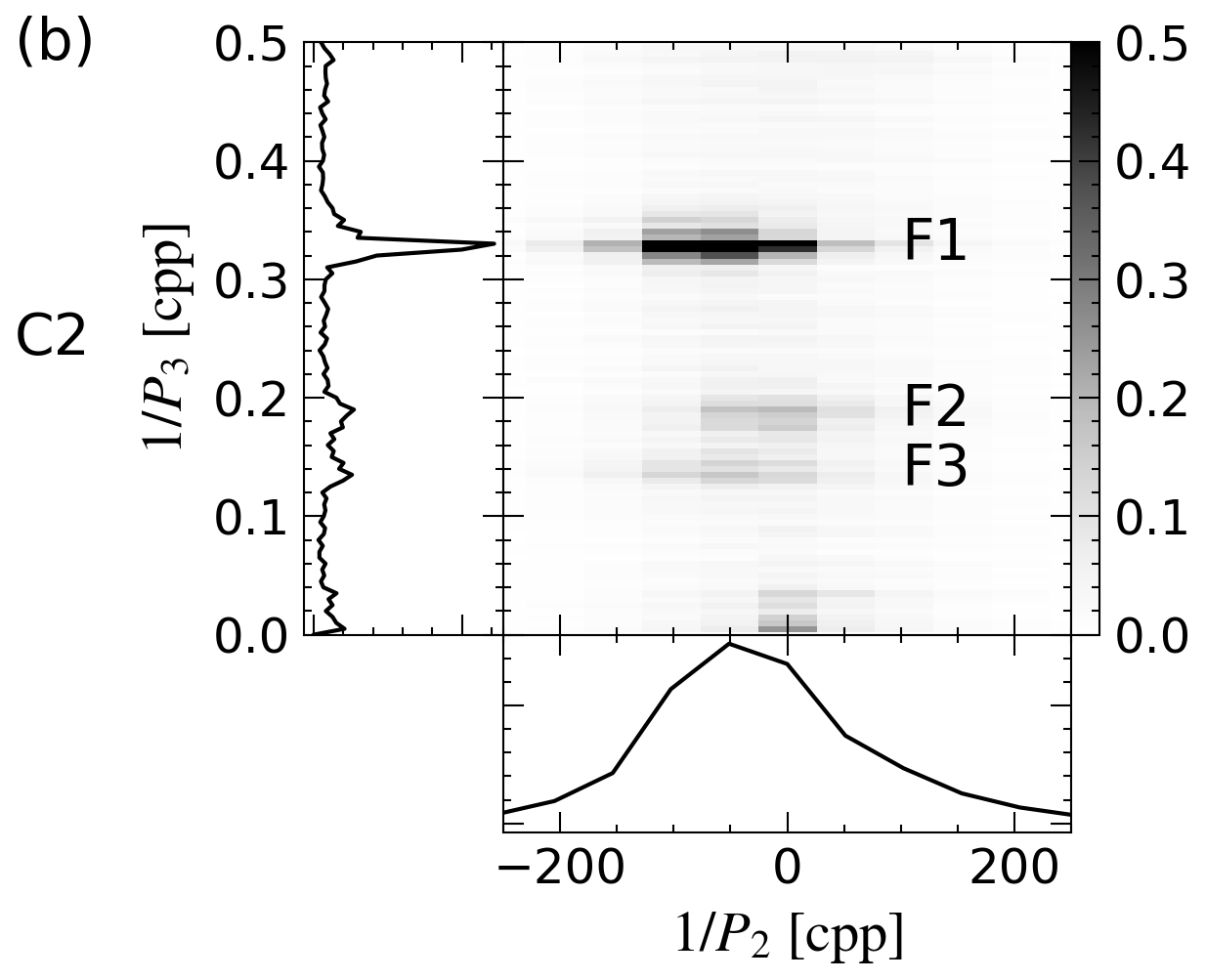}
\caption{The magnitude (square root of power) 2DFS of PSR~J1901$-$0906 for (a) the leading profile component (C1) and (b) the trailing profile component (C2) using the same data as in \fig{fig:LRFS} with an FFT length of 200 pulse periods. In each panel, the left and bottom side panels show the horizontally and vertically integrated power, respectively. Similar to \fig{fig:LRFS}, the DC row ($1/P_3 = 0$) has been set to zero and each spectrum is normalized separately with the grey scale capped at 50 per cent.
Three spectral features, F1, F2, and F3, are labelled.}
  \label{fig:2DFS}
\end{figure}

We computed longitude-resolved fluctuation spectra \citep[LRFS,][]{LRFS} and two-dimensional fluctuation spectra \citep[2DFS;][]{es02} for each component using {\sc psrsalsa}\footnote{\url{https://github.com/weltevrede/psrsalsa}} \citep{PSRSALSA}.
For this analysis, we used TPA observations longer than the FFT length of 200 pulse periods, from which continuous blocks of 200 pulsar rotations were extracted.
The resulting LRFS is presented in panel (b) of \fig{fig:LRFS}, while the 2DFS\footnote{As explained in \citetalias{hww+25}, by convention, all 2DFS presented in this paper are horizontally flipped compared to the mathematical definition of the 2D Fourier transform.} is shown in \fig{fig:2DFS}.
In both C1 and C2, three spectral features are identified, labelled F1, F2, and F3, in addition to a low-frequency spectral feature corresponding to the nulling, which will be discussed in \sect{sec:null}.

In C1, F2 and F3 are much stronger than F1, whereas F1 dominates in C2.
Consequently, the drifting subpulses in C2 as seen in the pulse stack (\fig{fig:pulsestack}) are primarily associated with F1, producing a more regular drift pattern.
In contrast, the comparable strength and similar $1/P_3$ frequencies of F2 and F3 in C1 give rise to a more complex-looking emission pattern.
Moreover, F2 becomes strong at a slightly later pulse longitude compared to F3, extending further into the weaker bridge region, as shown in the LRFS in panel (b) of \fig{fig:LRFS}.
The remaining panels of this figure will be discussed in \sect{sec:pol}.

The horizontal separation in pulse longitude ($P_2$) and vertical spacing in pulse number ($P_3$) for the spectral features (centroid positions in 2DFS) for both C1 and C2 are measured following the methods of \cite{wes06, wse07, sws+23} and summarized in \tab{tab:measure}.
The quoted uncertainties are representative estimates, combining the spectral resolution limits and the subjective identification of the feature positions.
The robustness of these measurements is verified by their consistency for independent halves of the data and the (not shown) time-resolved sliding two-dimensional fluctuation spectrum \citep[S2DFS, see][]{ssw09}.
Within the uncertainties, all spectral features have consistent $P_2$ and $P_3$ values in C1 and C2, and these $P_3$ values are also consistent with previous studies \citep{wes06, wse07, njm+17, bmm+16, bmm+19, sws+23}.
It is clear that F1, F2, and F3 are not harmonically related, as no feature has both its $1/P_2$ and $1/P_3$ frequencies equal to the same integer multiple of the corresponding frequencies of another feature\footnote{This conclusion remains valid even if one or more of F1, F2, and F3 are aliased, since aliasing only changes the sign of the corresponding $1/P_2$ frequency and does not change the magnitude of any ratio between their $1/P_2$ frequencies.}.
However, remarkably, their $1/P_2$ and $1/P_3$ frequencies (in units of cycles per period; cpp) do satisfy the relation $\rm F1=F2+F3$ precisely with any deviation well within the quoted errors.
This clearly demonstrates that the three periodicities are not independent, a conclusion which is strengthened in the next subsection, where it is shown that the associated subpulse phases are related via an identical relation.

\begin{table}
\caption{$1/P_3$ and $1/P_2$ measurements of the centroid positions of the three spectral features labelled in \fig{fig:2DFS}.}
\label{tab:measure}
\centering
\begin{tabular}{cccccc}
\toprule
  &  & $1/P_3$ & $P_3$ & $1/P_2$ & $P_2$ \\
  &  & (cpp) & & (cpp) & ($^\circ$) \\
\midrule
    & F1 & $0.328\pm0.006$  & $3.05\pm0.05$  & $-66\pm51$   & $-5\pm4$    \\ [0.7em]
C1  & F2 & $0.188\pm0.007$  & $5.31\pm0.21$  & $-7\pm51$    & -           \\ [0.7em]
    & F3 & $0.136\pm0.005$  & $7.33\pm0.27$  & $-50\pm51$   & $-7\pm7$    \\ [0.7em]
\midrule
    & F1 & $0.328\pm0.008$  & $3.05\pm0.08$  & $-68\pm51$   & $-5\pm4$    \\ [0.7em]
C2  & F2 & $0.186\pm0.009$  & $5.38\pm0.27$  & $-8\pm51$    & -           \\ [0.7em]
    & F3 & $0.140\pm0.006$  & $7.14\pm0.31$  & $-52\pm51$   & $-7\pm7$    \\ [0.7em]
\bottomrule
\end{tabular}
\end{table}

If PSR~J1901$-$0906 is a drift mode changer, a common and natural explanation for the presence of multiple periodicities, one would expect three distinct drift patterns corresponding to F1, F2 and F3 alternating in the pulse stack over time with abrupt transitions between different modes \citep[e.g.][]{htt70, rwr05, smk05}. 
However, at least in C2, where F1 dominates, F1 remains clearly visible throughout the pulse stacks in all observations as long as the pulsar is not in the null state. This rules out the interpretation of F1 being an intermittent drift mode.
Disentangling F2 and F3 in the pulse stacks is much more challenging due to their comparable $1/P_3$ frequencies.
If F2 and F3 correspond to two distinct drift modes, how these mode changes can be identified depends on the time-scale over which the modes alternate.
For long time-scales, say, tens of $P_3$, the modes will persist long enough to be discernible directly in the pulse stack as switches between well-defined modes. 
This is not observed in C1, which has an irregular appearance of the drifting subpulses in the pulse stack.
Conversely, if these modes alternate on a much shorter time-scale $\sim P_3$, the drifting subpulse pattern would be irregular. However, then one cannot expect to observe two well-defined spectral features in the Fourier domain; instead, the spectral power would be smeared along the $1/P_3$ direction.
Therefore, the irregular appearance of the drifting subpulses of C1, combined with well-defined and separated F2 and F3 features in the fluctuation spectra, strongly suggests F2 and F3 must co-exist in C1. Moreover, since F1 is persistently detected in C2, F2 and F3 must co-exist with F1.

\subsection{Subpulse phase and modulation amplitude}
\label{sec:subpulse_phase}

In \sect{sec:F123} we established that F1, F2, and F3 are not independent, as both their $1/P_2$ and $1/P_3$ frequencies obey the relation $\rm F1=F2+F3$. In what follows, we show that the subpulse phases associated with F1, F2, and F3 follow an identical relation.
We computed the subpulse phase and modulation amplitude for F1, F2, and F3, following \cite{es02, es03a, PSRSALSA}.
This analysis is done using all available TPA data, from which continuous blocks of 64 pulsar rotations were extracted. Only blocks for which $\ge60$ individual pulses exhibited a signal-to-noise ratio $\ge7$ were selected. This selection removed 29 of the 106 blocks, resulting in a total of 4928 pulses contributing to the analysis\footnote{We verified that all conclusions presented in this paper remain qualitatively unchanged when all available pulsar rotations, including null pulses, are included. However, the selection does improve the statistical significance of the analysis.}.

\begin{figure}
\centering
\includegraphics[width=\hsize]{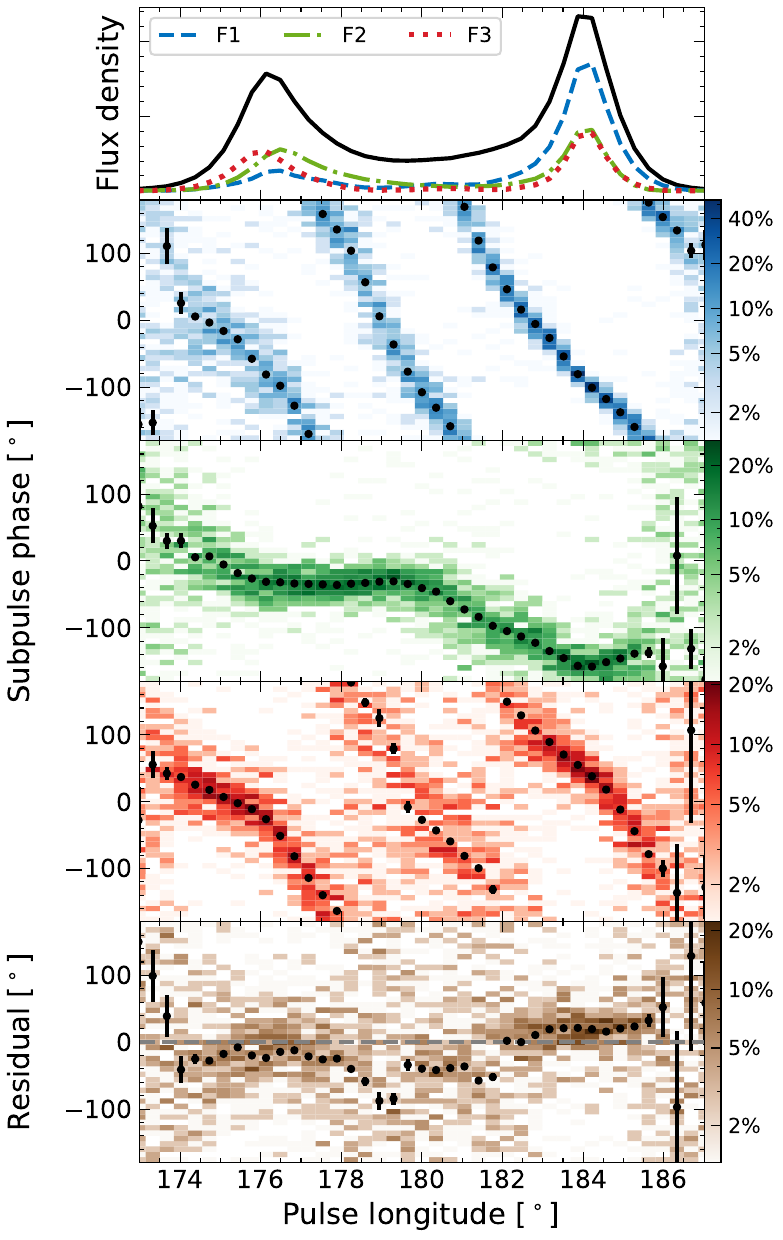}
  \caption{
  The top panel displays the pulse profile (solid black), and  the subpulse modulation amplitudes for F1 (dashed blue), F2 (dash–dotted green), and F3 (dotted red).
  In the rest of the panels, the distributions of the subpulse phases of F1 (blue), F2 (green), F3 (red), and the residuals relative to the expectation from a beat relation (brown, see main text for the definition) are displayed along with the block-averaged subpulse phases (black dots) and their error bars.
  The colour indicates the fraction of 64-pulse-long blocks showing a given subpulse phase at a given pulse longitude. 
  The horizontal dashed grey line in the bottom panel marks the zero residual.
  }
  \label{fig:subpulse_phase}
\end{figure}

The results of this analysis are shown in \fig{fig:subpulse_phase}. The top panel compares the longitude-resolved subpulse modulation amplitudes of F1, F2, and F3 (dashed blue, dash-dotted green, and dotted red lines, respectively) with the average pulse profile (solid black line).
These amplitudes are normalized such that a pure sinusoidal modulation would have the same amplitude as the average pulse profile.
The F1 amplitude (dashed blue) reaches its maximum in C2, whereas it is weaker than F2 and F3 in C1. This change of dominance of spectral features is consistent with Figs.~\ref{fig:LRFS} and \ref{fig:2DFS}.
As shown in \fig{fig:LRFS}, the peak of the F2 amplitude lags that of F3 by approximately $1^\circ$ in pulse longitude in C1, while no significant differences are observed in C2.

The three middle panels of \fig{fig:subpulse_phase} display the longitude-resolved subpulse phase distributions for F1 (blue), F2 (green), and F3 (red), from top to bottom, respectively. 
The colour maps display the histograms of subpulse phases accumulated from all 64-pulse-long blocks at each pulse longitude bin.
These subpulse phases from different blocks have been aligned using the algorithm defined in \cite{es03a}.
The overlaid black dots indicate the block-averaged subpulse phases, with error bars determined via bootstrap as described in \cite{PSRSALSA}.
Overall, the error bars reflect the strength of the pulsar emission relative to the system noise level, while the width of the distribution in the colour map characterizes the intrinsic stability of the spectral feature across different blocks.

The subpulse phase tracks of F1 and F3 exhibit similar negative gradients\footnote{Consistent with the convention adopted in \fig{fig:2DFS}, the subpulse phases here differ by a sign flip from the conventional Fourier definition (see \citetalias{hww+25} for more details).}, corresponding to a $1/P_2$ value of about $-66$~cpp, as listed in \tab{tab:measure}. The sign convention is such that this corresponds to negative drift: drift from the trailing to leading side of the pulse profile as seen in \fig{fig:pulsestack}.
The absence of a subpulse phase jump for F1, such as those observed in some pulsars \citep[see, e.g.,][]{es03a, wkw+25}, indicates that the drifting subpulses with $P_3 \approx 3.03$ extend smoothly across both profile components.
This can be confirmed through careful visual inspection of the pulse stacks, such as shown in \fig{fig:pulsestack}, including in the weaker bridge region ($178^\circ\sim182^\circ$) between C1 and C2.
For F3, there seems to be a small phase jump around a pulse longitude of $\approx179.5^\circ$. However, given the very low modulation amplitude and the relatively large subpulse phase instability across blocks, it is difficult to judge its significance.

Between pulse longitudes $\approx176^\circ$ and $179^\circ$, where F2 is the dominant spectral feature, its subpulse phase track has a zero gradient, corresponding to amplitude modulation, in contrast to the negative gradients of F1 and F3.
In addition, the subpulse phase track of F2 has kinks such that the leading parts of both C1 and C2 show negative gradients.
However, the trailing part of C2 possibly suggests the presence of a positive drift. This reversal of the drift direction across the pulse profile components is known as bi-drifting \cite[e.g.,][]{PSRSALSA, wkw+25}. The transition longitude seems to coincide with the peak of the pulse profile in C2. As the emission drops sharply toward the trailing edge of C2, it remains to be seen how significant this reversal of drift direction is.

If the subpulse modulation satisfies the beat relation $\rm F1=F2+F3$, the subpulse phases should follow the same relation \citepalias{hww+25}. If so, the phase residual $\rm F1-(F2+F3)$ is expected to be near zero.
This calculation is performed for all blocks, yielding a distribution of residuals (brown colour map) and the block-averaged residual (black dots) as presented in the bottom panel of \fig{fig:subpulse_phase}.
If the subpulse phases for F1, F2, and F3 were independent and varied stochastically across all blocks, a uniform (random) distribution of the residuals would be expected. In contrast, the observed residuals cluster around zero, particularly at pulse longitudes where the modulation amplitudes are large and the measured subpulse phases are most accurate.
Our results prove that not only $1/P_2$ and $1/P_3$ frequencies, but also the subpulse phases adhere to the relation $\rm F1=F2+F3$.
This unambiguously demonstrates that the three fluctuation frequencies are intimately related, and cannot be generated by three separate independent physical processes.

\subsection{Beat system}
\label{sec:beat}
The relation $\rm F1=F2+F3$ identified in both the $1/P_2$ and $1/P_3$ frequencies (\sect{sec:F123}) as well as the subpulse phases (\sect{sec:subpulse_phase}) for the co-existing periodicities is reminiscent of the subpulse modulation in PSR~J1514$-$4834 \citepalias{hww+25}. This strongly suggests that a ‘beat system’ plays a role, in the sense that two co-existing underlying fluctuation frequencies modulate each other. Such a modulation would result in two additional ‘beat features’, corresponding to the sum and difference of the two fluctuation features which modulate each other. As explained and demonstrated in \citetalias{hww+25}, the sum and difference apply to the $1/P_2$ and $1/P_3$ frequencies as well as the subpulse phases. The established relation $\rm F1=F2+F3$ for PSR~J1901$-$0906 would correspond to one of these.

As a consequence, if a beat system operates in this pulsar, at least one additional beat feature is to be expected.
In the case of PSR~J1514$-$4834, it has been shown that even when a beat feature cannot be clearly identified in a power spectrum, there is the potential for a phase spectrum to reveal its presence.
Therefore, in what follows, we adopt the methodology of \citetalias{hww+25} and explore phase spectra to not only confirm that F1, F2 and F3 follow the expectations of a beat system, but, more importantly, establish the existence of the expected additional beat features. This includes beat features arising from harmonics.

\begin{figure*}
    \raggedright
    \includegraphics[width=0.994\linewidth]{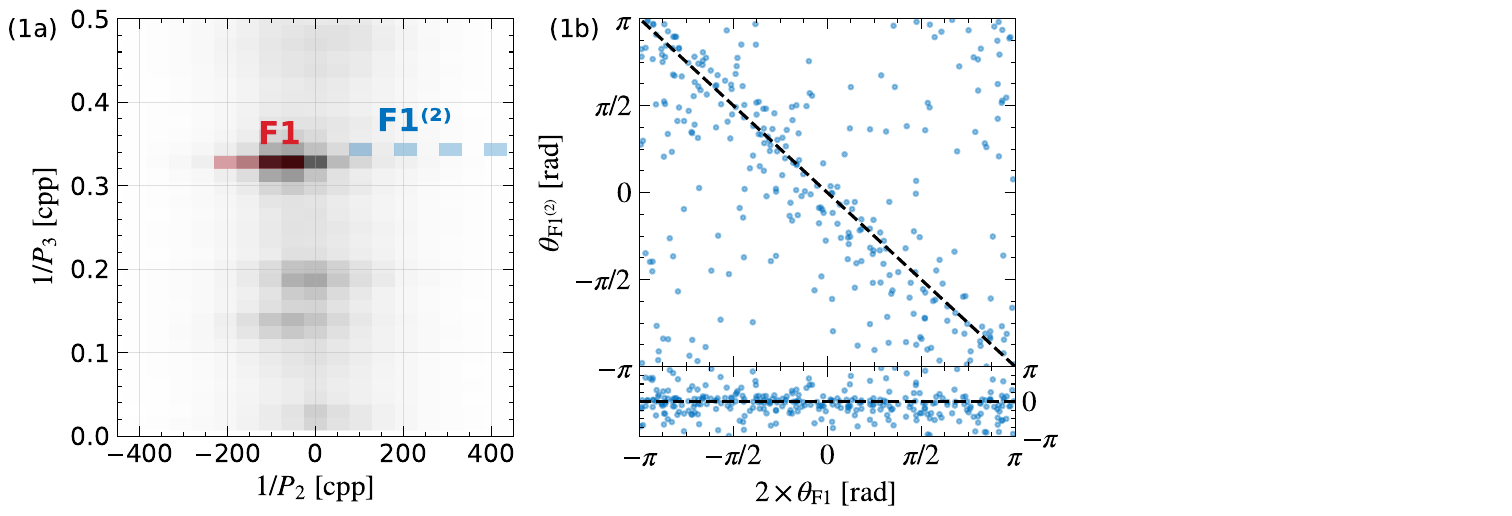} \\
    \includegraphics[width=\linewidth]{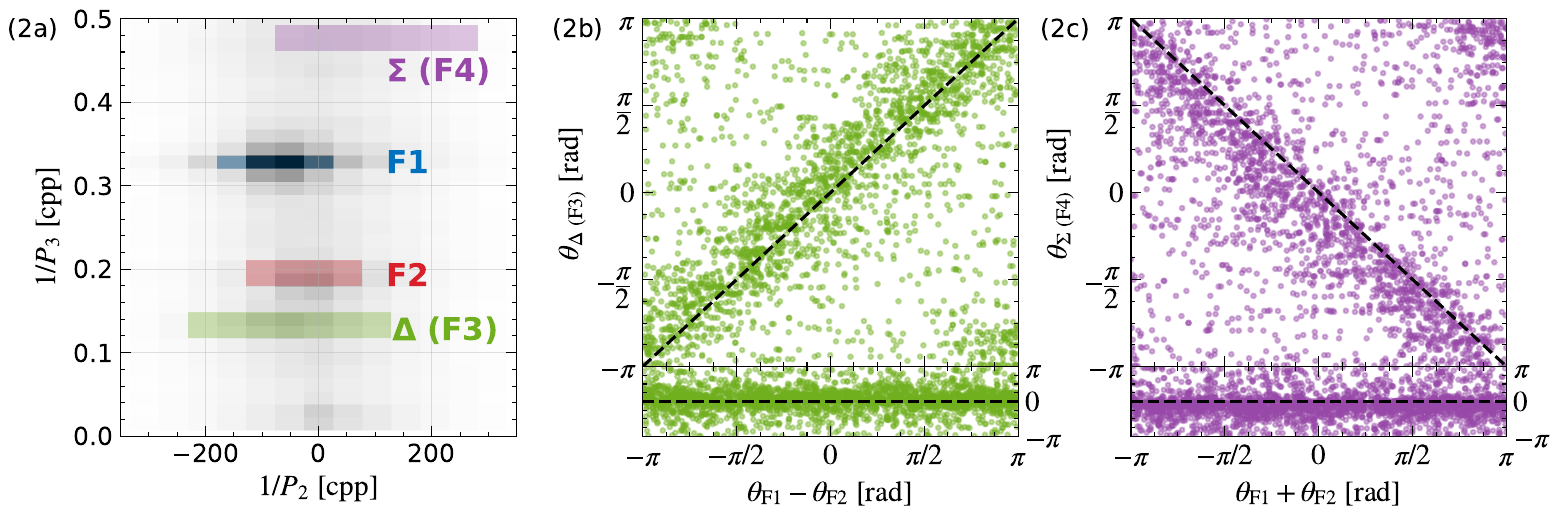} \\
    \includegraphics[width=\linewidth]{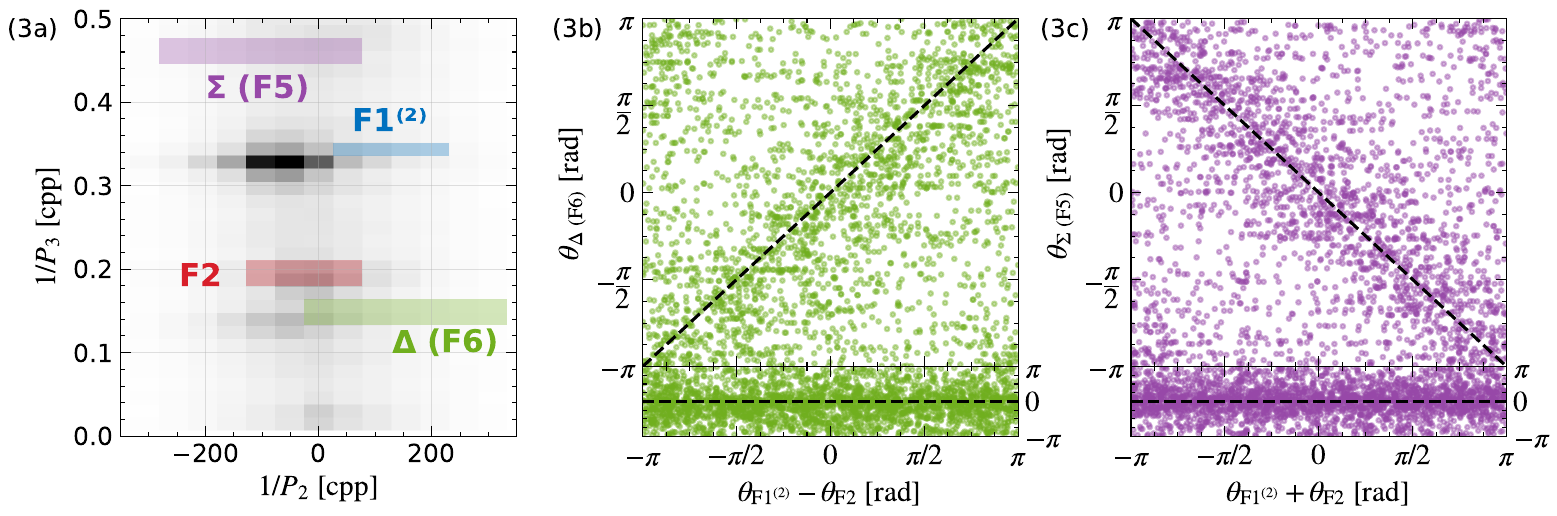}
    \caption{
    \textbf{(1a):} Magnitude (square root of power) 2DFS of C2 computed from the same data as shown in \fig{fig:subpulse_phase} with an FFT length of 64 pulse periods. Four pixels from the fundamental feature F1 are highlighted in the red shaded region, and their (aliased) second harmonics $\rm F1^{(2)}$ are indicated by the blue shaded region.
    \textbf{(1b):} Scatter plot of the predicted Fourier phases of the second harmonic (i.e., twice the Fourier phase of the fundamental frequency, $2\theta_{\rm F1}$) versus the measured phases, $\theta_{\rm F1^{(2)}}$. 
    The dashed diagonal line indicates the expected correlation if the second harmonic is significantly detected, consistent with the measured circular correlation coefficient of $r=-0.30$.
    \textbf{(2a):} The same 2DFS as in (1a), but highlighting pixels from F1 (blue) and F2 (red) which are assumed to be modulating each other and paired to determine the predicted $\Delta$ (F3, green) and $\Sigma$ (F4, aliased, purple) beat features.
    \textbf{(2b, 2c):} Scatter plots similar to (1b), but comparing the predicted Fourier phases of the beat features with the measured phases.
    The circular correlation coefficients are $r_{\Delta~(\rm F3)}=0.34$ (panel 2b) and $r_{\Sigma~(\rm F4)}=-0.28$ (panel 2c).
    \textbf{(3a--3c):} Similar to (2a--2c) but with $\rm F1^{(2)}$ and F2 modulating each other. The circular correlation coefficients are $r_{\Delta~(\rm F6)}=0.11$ (panel 3b) and $r_{\Sigma~(\rm F5)}=-0.15$ (aliased, panel 3c). 
    In all scatter plots, the bottom sub-panels show phase residuals relative to the expected correlation: $y-x$ in panels (2b) and (3b), and $y+x$ in panels (1b), (2c), and (3c), where $x$ and $y$ denote the phases plotted on the horizontal and vertical axes, respectively. The plus sign accounts for the phase reversal caused by aliasing. The dashed horizontal lines indicate zero residual.
    The analysis was performed across 77 independent data blocks, meaning each unique pixel combination contributes 77 data points in the scatter plots. Consequently, (1b) displays 308 points (from four F1 pixels), while the scatter plots in (2b, 2c) and (3b, 3c) each display 2464 data points (from 32 pixel combinations of four F1 or $\rm F1^{(2)}$ pixels paired with eight F2 pixels).
    }
    \label{fig:scatter-plot-F1-F2}
\end{figure*}

Before searching for expected beat features, it is informative to first identify the second harmonic of F1 in C2. This harmonic is labelled $\rm F1^{(2)}$, and, although weak, will be shown to play a crucial role in explaining the complexities in the spectra of PSR~J1901$-$0906.
The second harmonic $\rm F1^{(2)}$ has intrinsic $1/P_2$ and $1/P_3$ frequencies that are twice those of F1. Its Fourier phase is also expected to be twice that of F1 (modulo $2\pi$), given the form of the subpulse modulation considered (see \apdx{apdx:harmonic_phase}).
Due to the aliasing effect, this spectral feature should emerge at $1/P_2\approx136~\rm cpp$ and $1/P_3\approx0.344~\rm cpp$ in the 2DFS, as shown in blue in panel (1a) of \fig{fig:scatter-plot-F1-F2}.
Since the power of higher harmonics decreases with increasing order, 
the power of this feature is comparable to the level of random fluctuations in the pulsar emission, making it indistinguishable in the power spectrum.
However, its presence is confirmed by the strong correlation exhibited in the Fourier phase analysis as shown in panel (1b) of \fig{fig:scatter-plot-F1-F2}, where a negative rather than a positive correlation arises from the aliasing effect (see \citetalias{hww+25} for the definition of the circular correlation coefficient and further details on aliasing).
The statistical significance of all phase correlations is quantified in \apdx{apdx:additional_analysis}.

If F1 and F2 modulate each other (this is not necessarily the only interpretation, as will be discussed below and in \apdx{apdx:beat_harmonic_alias}), a corresponding modulation of the second harmonic $\rm F1^{(2)}$ and F2 is naturally expected. With both F1 and $\rm F1^{(2)}$ being detected, not only beats between F1 and F2, but also those between $\rm F1^{(2)}$ and F2, are potentially detectable.
The former are explored in the second row of \fig{fig:scatter-plot-F1-F2}, while the latter are presented in the third row, both of which are discussed individually below.

In panel (2a), the power in F1 and F2, which are assumed to modulate each other, is spread out in the 2DFS. Therefore, four pixels corresponding to F1 (highlighted in blue) and eight pixels corresponding to F2 (indicated in red) are considered.
Different pixel combinations (one of the four pixels from F1 paired with one of the eight pixels from F2) result in different expected positions for the $\Sigma$ (the sum) and $\Delta$ (the difference) beat features, as indicated in purple and green respectively.
In this interpretation, the $\Delta$ beat feature would correspond to F3, while the $\Sigma$ beat feature would be observed to be aliased and would correspond to an additional expected feature (labelled F4).
As shown in the middle panel (2b), the measured Fourier phases of F3 (the $\Delta$ beat) are entirely predictable given the observed Fourier phases of F1 and F2, resulting in a strong correlation observed in the scatter plot.
This result is expected given the already established relation $\rm F3=F1-F2$.
This is because, as discussed in \citetalias{hww+25}, the strong correlation observed in the Fourier phases in \fig{fig:scatter-plot-F1-F2} is mathematically equivalent to the constant residuals found in the subpulse phases (bottom panel of \fig{fig:subpulse_phase}).
More interesting is the identification of the $\Sigma$ beat feature (F4).
This feature, although undetectable in the power spectrum, is confirmed by the strong correlation exhibited in the right-hand scatter plot (2c). Therefore, indeed two beat features are detected, as expected when two frequencies modulate each other.

We repeat the same procedure to verify the presence of beat features between $\rm F1^{(2)}$ and F2, with the results presented in the third row of \fig{fig:scatter-plot-F1-F2}.
Both the $\Sigma$ (F5) and $\Delta$ (F6) beat features are detected in the Fourier phase correlation plots (panels 3b and 3c), with the former being aliased.
The significance of the correlations is lower given the weaker power of $\rm F1^{(2)}$ compared to F1 (see \apdx{apdx:additional_analysis} for a quantitative analysis).

\begin{figure*}
    \centering
    \includegraphics[width=\linewidth]{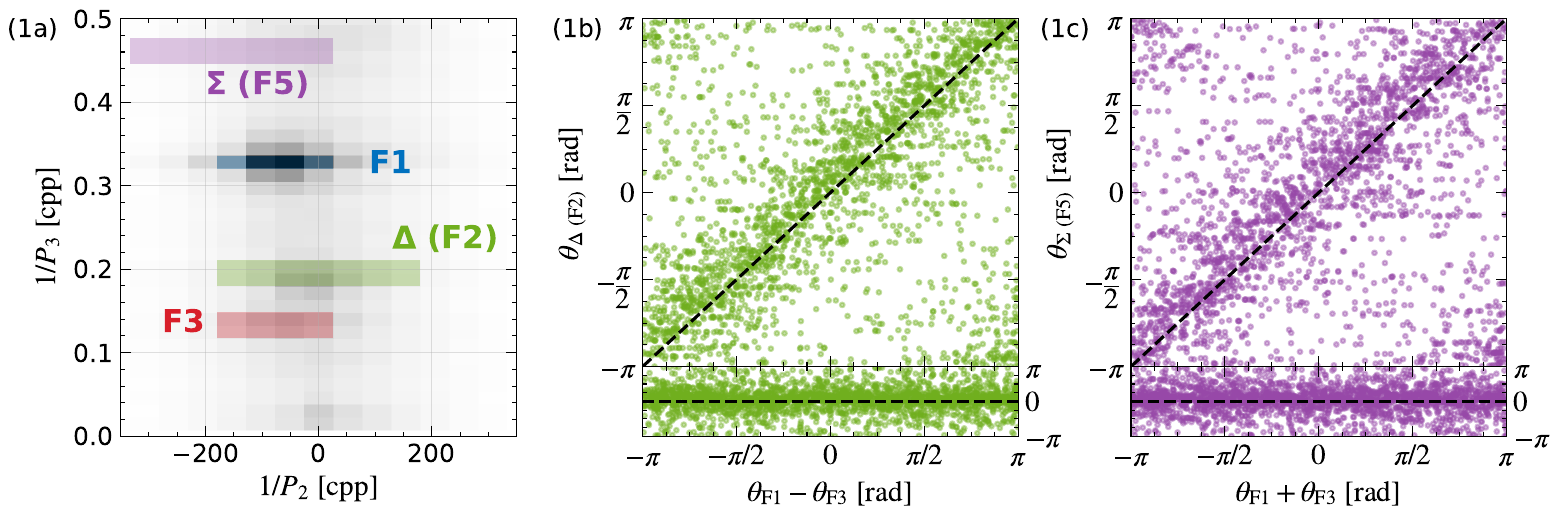} \\
    \includegraphics[width=\linewidth]{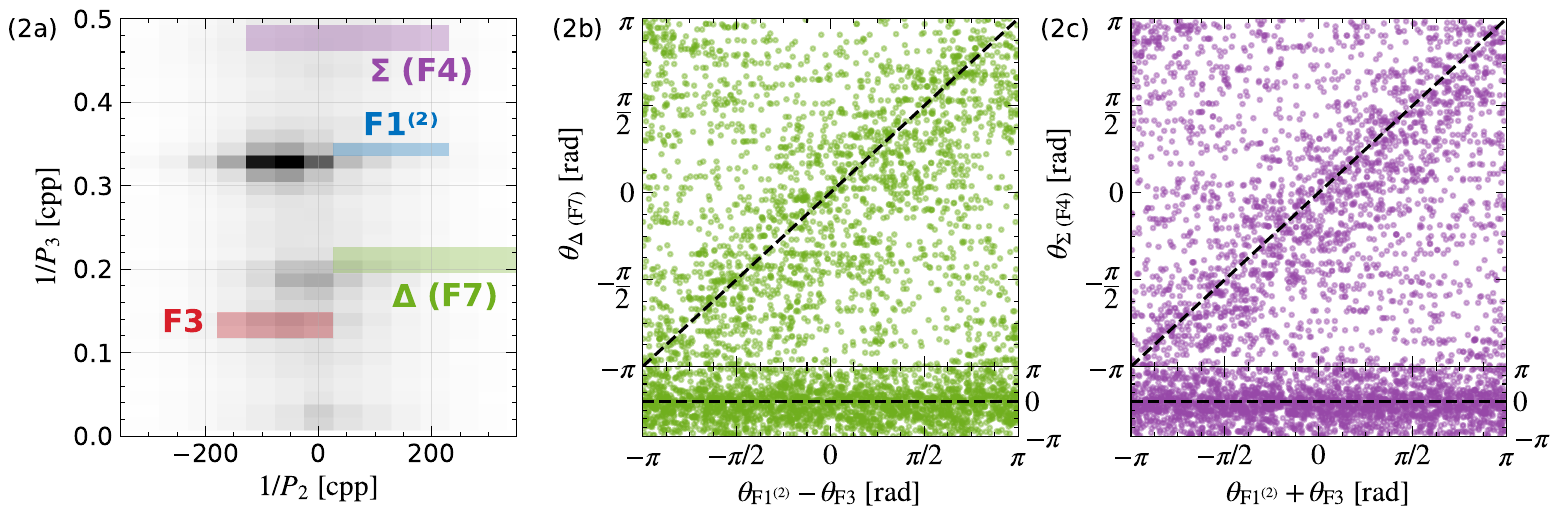} \\
    \caption{
    Similar to \fig{fig:scatter-plot-F1-F2}, but assuming that F3 and either F1 (1a--1c), or its second harmonic $\rm F1^{(2)}$ (2a--2c), modulate each other in C2.
    With 32 possible pixel combinations (four from F1 or $\rm F1^{(2)}$ paired with eight from F3), each scatter plot displays a total of 2464 data points.
    The circular correlation coefficients are $r_{\Delta~(\rm F2)}=0.35$ (1b), $r_{\Sigma~(\rm F5)}=0.29$ (1c), $r_{\Delta~(\rm F7)}=0.10$ (2b) and $r_{\Sigma~(\rm F4)}=0.11$ (2c).
    }
    \label{fig:scatter-plot-F1-F3}
\end{figure*}

Our Fourier phase analysis in \fig{fig:scatter-plot-F1-F2} suggests that F1 (and its harmonic) and F2 modulate each other in C2, resulting in the observable beats. 
However, starting from
the relation $\rm F1=F2+F3$, this is not a unique interpretation.
Alternatively, \fig{fig:scatter-plot-F1-F3} explores a different interpretation where F1 (and its harmonic) and F3 modulate each other in C2 instead.
As shown in panel (1a), in this interpretation, F2 would be the $\Delta$ beat feature between F1 and F3, showing a significant correlation in panel (1b), which is further quantified in \apdx{apdx:additional_analysis}.
Distinguishing between F1 being modulated by F2 or F3 is hard because these two interpretations are mathematically equivalent and identical beat features would be expected (see \apdx{apdx:beat_harmonic_alias} for a detailed derivation and examples).
More generally, any beat features arising from the modulation between F1 (and its harmonics) and F2 can be equivalently expressed as beat features between F1 (and its harmonics) and F3.
Consequently, all spectral features in C2 can be explained by F1 (along with its harmonics) being modulated by either F2 or F3.
Regardless of whether F2 or F3 is treated as the fundamental modulation frequency of F1, it is remarkable that the plethora of spectral features and their phase correlations can be reduced to essentially two fundamental frequencies, their harmonics, and their associated beat frequencies\footnote{This conclusion holds regardless of whether F1, F2, and F3 themselves are already aliased (see \apdx{apdx:beat_harmonic_alias} for a detailed discussion).}.
This suggests the model proposed for PSR~J1514$-$4834, which was based on two fundamental frequencies modulating each other, can be applied here as well.
This model is discussed in \sect{sec:feedback}.

In addition to the two aforementioned interpretations: beats between F1 and F2 (\fig{fig:scatter-plot-F1-F2}) and beats between F1 and F3 (\fig{fig:scatter-plot-F1-F3}), a third possibility is that F2 and F3 modulate each other in C2, resulting in F1 as a $\Sigma$ feature.
However, there is no detectable $\Delta$ beat feature between F2 and F3, thereby disfavouring this interpretation (see \apdx{apdx:additional_analysis}).

In C1, where F2 and F3 are strong and of comparable strength and unlike for C2, there is no evidence for additional beat features arising from combinations of F1, F2, and F3 (see Appendices~\ref{apdx:scatter-plot-F2-F3} and \ref{apdx:additional_analysis}). Hence, despite the relation $\rm F1=F2+F3$ being obeyed, it is unclear if this is the result of two fluctuation frequencies modulating each other in C1.

\section{Polarization properties}
\label{sec:pol}

\begin{figure}
    \centering
    \includegraphics[width=\linewidth]{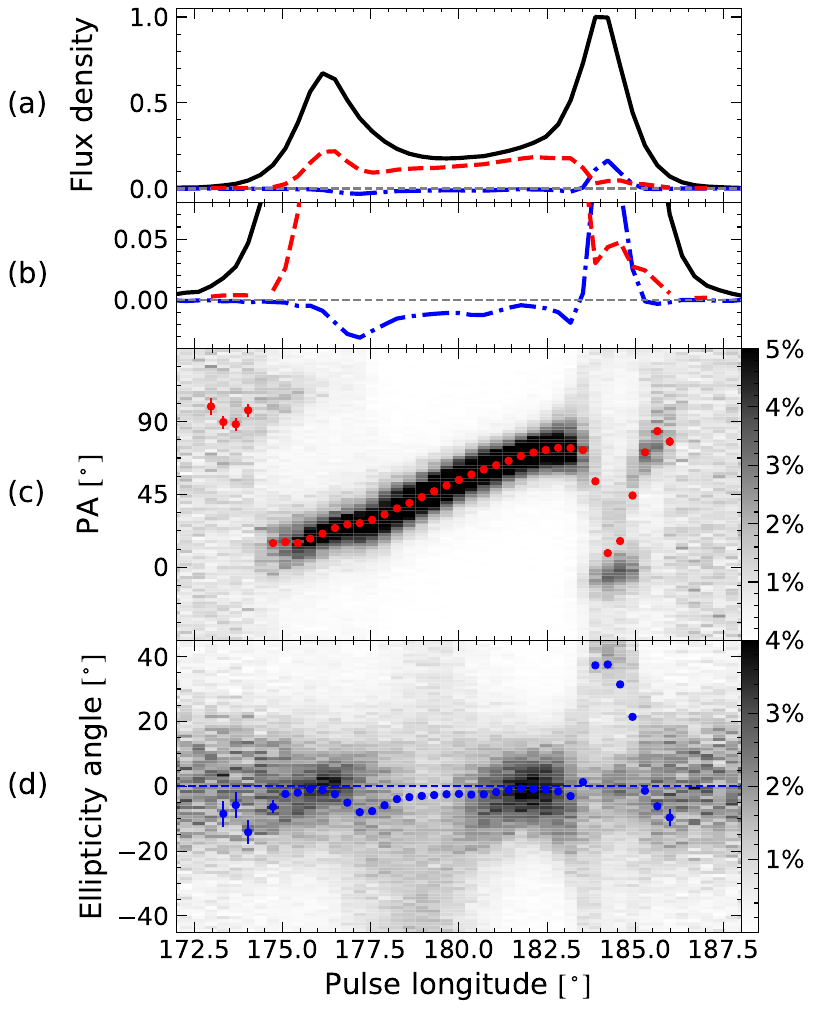}
    \caption{
    (a): Normalized average total intensity pulse profile of PSR~J1901$-$0906 ($I$, solid black line), linear polarization intensity ($L$, dashed red line), and circular polarization intensity ($V$, dash-dotted blue line). All profiles are normalized to the peak amplitude in total intensity.
    (b): A zoomed-in view of panel (a).
    (c): The linear polarization position angle (PA) distribution obtained from samples from individual pulses where the significance of $L$ exceeds $2\sigma$ (null state pulses are excluded). The grey scale indicates the fraction of pulses exhibiting a specific PA at a given pulse longitude, with a dynamic range chosen such that black corresponds to 5 per cent or more.
    The PA obtained from the average profile is superimposed as red points with error bars.
    (d): Similar to panel (c), but for the ellipticity angle distribution, with the grey scale capped at 4 per cent.
    Only samples with $\sqrt{Q^2+U^2+V^2} > 2\sigma$ are included.
    The horizontal dashed line indicates an ellipticity angle of $0^\circ$.
    }
    \label{fig:PA-EA}
\end{figure}

We have detailed the complex modulation properties of the total intensity radio emission of PSR~J1901$-$0906 in \sect{sec:J1901}.
To obtain further insight into how these modulations interact, we explore their polarization properties.
The average polarization profiles of PSR~J1901$-$0906 are presented in panels (a) and (b) of \fig{fig:PA-EA}.
Following \cite{wk74, ew01, PSRSALSA}, the linear polarization intensity $L$ is corrected by bias-subtracting the noise contribution of Stokes $Q$ and $U$ measured in the off-pulse region. For bins where the pulsar signal is weaker than three times the noise level ($<3\sigma$), $L$ is undefined and is omitted from the figure.
At pulse longitude $\approx 183.5^\circ$, a sharp decrease in $L$ and a distinct surge in circular polarization $V$ are observed, which contrasts sharply with the slightly negative $V$ observed across the remainder of the pulse profile.

Panel (c) displays the phase-resolved distributions of the linear polarization position angle (PA).
Across C1 and the leading edge of C2, the PA follows a well-defined, monotonically increasing track.
This primary PA track, corresponding to the dominant linear polarization mode, is interrupted by an abrupt jump of nearly $90^\circ$, which coincides with the sharp decrease in $L$ and a prominent increase in $V$.
While the sharp decrease in $L$ is traditionally attributed to the incoherent addition of competing OPMs \citep[e.g.,][]{mth75, brc76}, the associated generation of $V$ could point to a partially coherent addition of these modes \citep{okj23}.
Interestingly, between pulse longitudes $\approx 183.5^\circ$ and $186.5^\circ$, the two preferred PA values are not strictly separated by $90^\circ$, as would be expected for ideal OPMs. Instead, the primary PA track splits into an upward and a downward branch at pulse longitude $\approx 183.5^\circ$. These branches connect to the secondary PA track and subsequently converge, tracing out a diamond-shaped structure.
OPM switching is also observed in the very leading edge of C1, where a secondary OPM is observed at PA $\approx 100^\circ$, representing an orthogonal offset compared to the dominant PA of $\approx 10^\circ$. In the following analysis, we establish this OPM switching is synchronized with at least F3.

The distribution of ellipticity angle in panel (d) features two clusters around $0^\circ$ at pulse longitudes $\approx176^\circ$ and $\approx182^\circ$, both of which correspond to the dominant linear polarization mode observed in panel (c). There is a hint of a structure resembling a ‘bow-tie’, such that in between these clusters the ellipticity angle flares out before converging again. After pulse longitude $\approx182^\circ$ the distribution splits again into an upward and a downward branch, before converging at pulse longitude $\approx185^\circ$. This ‘bow-tie’ shape is reminiscent of that observed in PSR~B1451$-$68 \citep{dwi21}. In addition, there is an additional cluster at $0^\circ$ ellipticity observed at pulse longitude $\approx185^\circ$, coincident with the position of the non-dominant OPM.

The LRFS based on the fractional linear polarization intensity ($L/I$) and the fractional circular polarization intensity ($|V|/I$) are presented in panels (c) and (d) of \fig{fig:LRFS}. Both show fluctuation features with the same periodicities as seen in the Stokes $I$ LRFS in panel (b), although the features, in especially panel (d), are less pronounced due to the weakness of $V$.
It is clear that F1, F2, and F3 do not only modulate the total intensity, but also modulate the polarization state.

Given that the polarization state is modulated by the F1, F2, and F3 cycles, we explore these cycles further by applying the $P_3$-folding technique.
Following \citet{iwj+20}, the Stokes $I$ pulse stack is first divided into 16-pulse blocks, and each block is folded at the $P_3$ period of F1, F2, or F3.
To compensate for variability in the $P_3$, cross-correlation is used to align the data between different blocks, thereby determining their relative phase offsets. 
These offsets are then applied when folding the Stokes $Q$, $U$, and $V$ pulse stacks.
Therefore, if features remain visible in the polarization $P_3$-folds after applying the offsets derived from Stokes $I$, it confirms that the polarization and total intensity modulations are synchronized.

Although most previous studies employed similar $P_3$-folding techniques to investigate the synchronization between total intensity and polarization, caution must be exercised when interpreting such results, especially for this pulsar with multiple co-existing periodicities.
The $P_3$-folding technique aligns any structures it finds during the cross-correlation alignment between blocks.
This is particularly concerning when folding with non-dominant periodicities, such as F2 and F3 in C2, causing structures associated with the dominant F1 cycle to show up in the $P_3$-folds for F2 or F3.
To address this, before the $P_3$-folding, we first isolated each spectral feature in the Fourier domain and performed an inverse Fourier transform to reconstruct pulse stacks containing only a range of periodicities of interest.
Considering the significant $P_3$ variability, the applied spectral filters need to be relatively broad.
We further verified the reliability of our results in two ways. 
Firstly, folding a spectrally isolated signal at an unrelated periodicity (e.g., folding the isolated F1 signal at the period of F2) yielded featureless noise. 
Secondly, we confirmed that the $P_3$-folds aligned by C1 alone, C2 alone, and the combined C1 and C2 data were qualitatively consistent.
These confirm that the identified modulations in our $P_3$-folds are physical in origin and not artifacts of the analysis method.

\begin{figure*}
    \centering
    \includegraphics[width=0.33\linewidth]{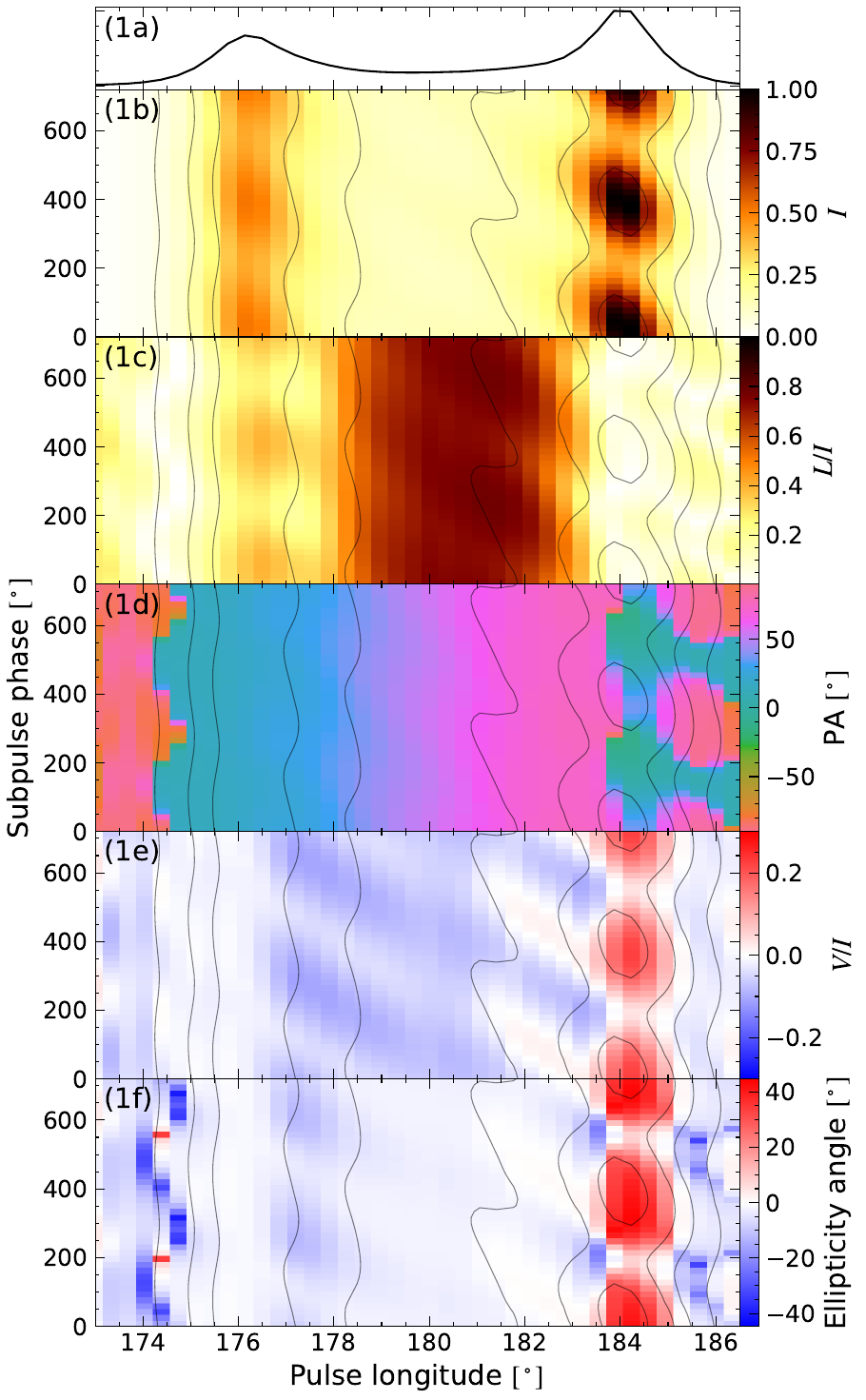}
    \includegraphics[width=0.33\linewidth]{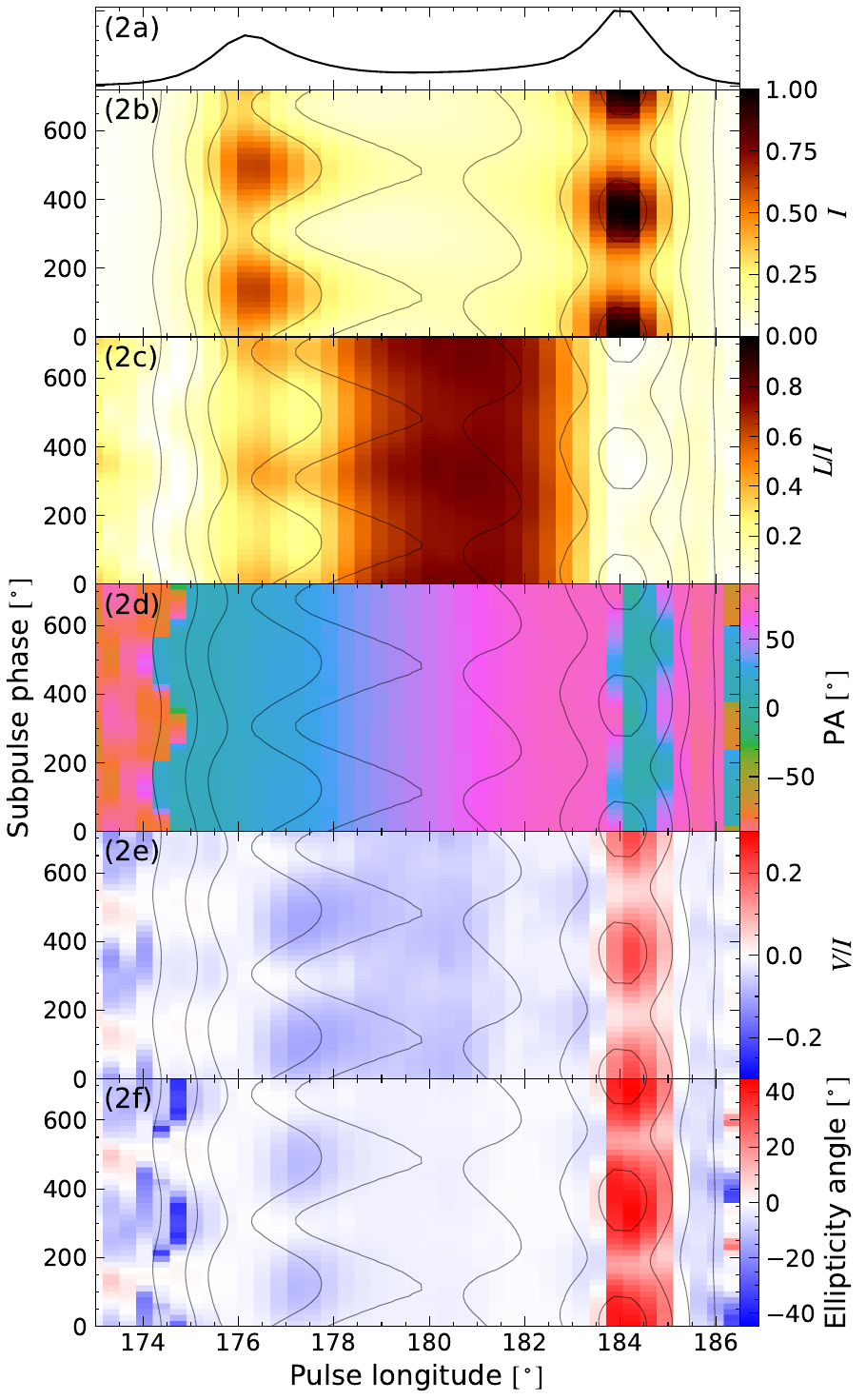}
    \includegraphics[width=0.33\linewidth]{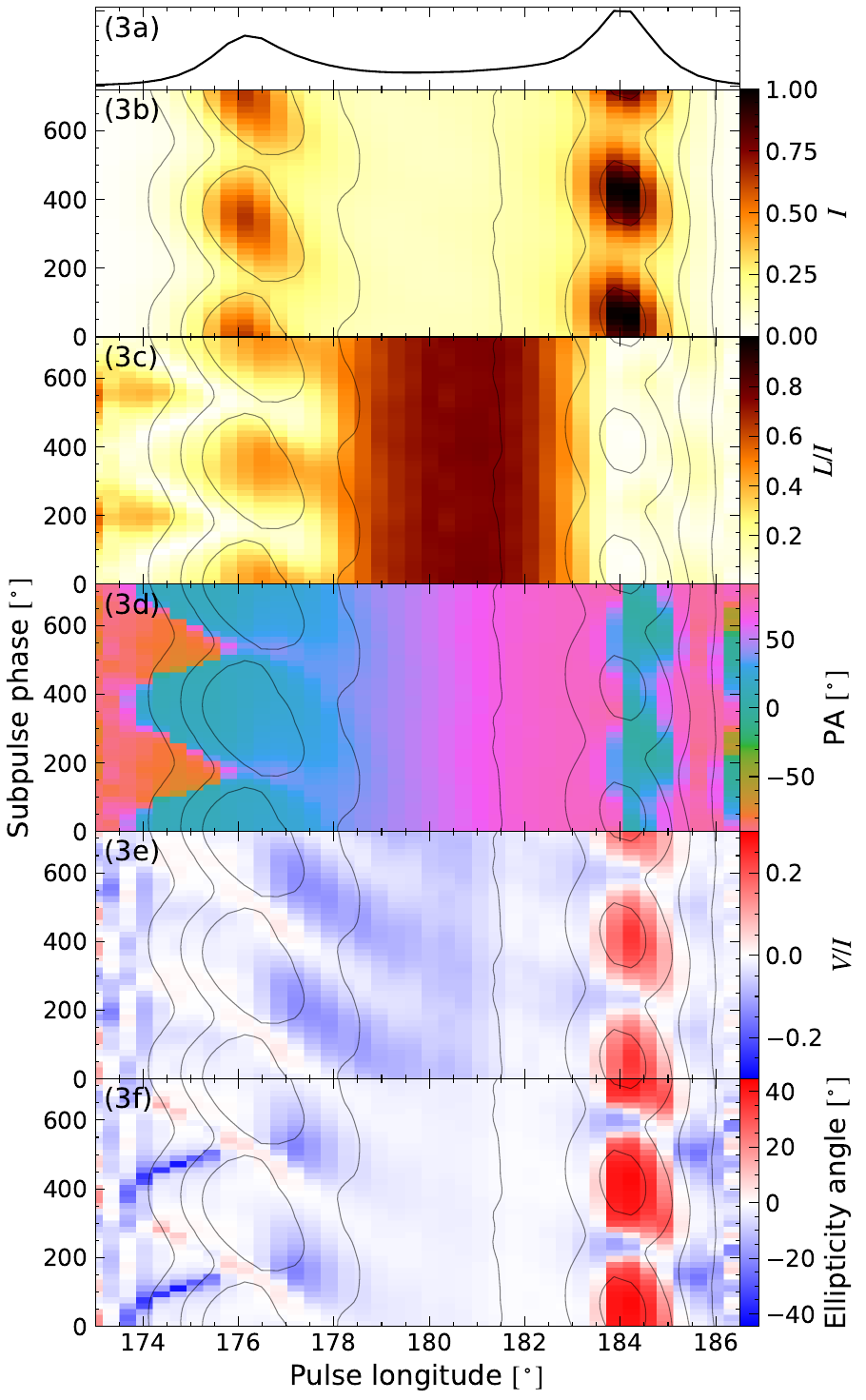}
    \caption{$P_3$-folds for PSR~J1901$-$0906, folded at F1 (left), F2 (middle), and F3 (right) after spectral filtering (see main text). The panels display, from top to bottom: (a) the pulse profile, the $P_3$-folds correspond to (b) $I$ (normalized to unity), (c) $L/I$, (d) PA, (e) $V/I$ (clipped at $\pm$0.3), and (f) ellipticity angle. For visual continuity, the modulation cycle is shown twice.
    The contours, identical across all panels, indicate the shape of the drift bands observed in panel (b).
    }
    \label{fig:p3fold}
\end{figure*}

The resulting $P_3$-folds for F1, F2, and F3, displaying Stokes $I$, fractional linear polarization intensity $L/I$, PA, fractional circular polarization intensity $V/I$, and ellipticity angle, are presented in \fig{fig:p3fold}.
For visual continuity, the average (folded) drift bands are plotted twice in each panel, covering two full $P_3$ cycles (subpulse phases from $0^\circ$ to $720^\circ$).
The contours in each panel correspond to the $P_3$-fold for Stokes $I$.
Previously identified polarization properties, including the low $L/I$, the OPM transition, and the positive $V$ in C2, are recovered in panels (c), (d) and (e) of \fig{fig:p3fold}, respectively. Moreover, the monotonically increasing PA track bridging C1 and C2 corresponds to a colour gradient from left to right in panel (d).

At the leading edge of C1 ($\approx175^\circ$) where F3 is strongest, the PA periodically oscillates between two OPMs (pink and cyan) in panel (3d). This confirms the bimodal nature of the PA and demonstrates that OPM switching is synchronized with F3.
As a consequence of this OPM mixing, a decrease of the $L/I$ is expected and observed in panel (3c).
This bimodal switching implies that the polarization state in the Poincar\'e space ($Q/I$, $U/I$, $V/I$) oscillates along a roughly linear trajectory passing through the origin within the $Q$-$U$ plane.

At the leading edge of C2 ($\approx182^\circ$), where F1 dominates and $V$ is relatively strong, the periodic modulation of $V/I$ and the ellipticity angle in panels (1e) and (1f), together with the constant PA in panel (1d), suggests that the oscillation of the polarization state follows a closed path confined to a plane orthogonal to the $Q$-$U$ plane.
Moreover, panel (1c) shows a decrease in $L/I$ as pulse longitude increases.
This implies that the centre of the oscillation path migrates toward the origin of the Poincar\'e space, eventually crossing the origin of the $Q$–$U$ plane.
Consequently, a PA jump (from pink to cyan) is observed at pulse longitude $\approx 183.5^\circ$ as shown in panel (1d).

The $P_3$-folds for F2 (middle column in \fig{fig:p3fold}) are similar to those of F1 and F3. However, unlike for F1 and F3, $L/I$ is high in between the Stokes $I$ drift bands in C1 (compare panels (2b) and (2c)).
This is also observed in PSR~B0031$-$07 \citep[drift mode B;][]{iwj+20}.

The $P_3$-folds in \fig{fig:p3fold} confirm that the overall polarization state exhibits periodic modulations that are synchronized with F1, F2, and F3 as observed in Stokes $I$.
We note that qualitatively similar conclusions can be drawn by applying the alternative method described in \cite{es04, edw04, ptv+22} which analyses the ‘polarization LRFS’ (PLRFS).

\section{Nulling}
\label{sec:null}

In addition to the co-existing periodicities and their associated beat features, PSR~J1901$-$0906 also shows intermittent nulling.
This is observed in the pulse stack (\fig{fig:pulsestack}), where two short-duration nulling intervals occur in close succession around pulse numbers 790 and 800.
Such intermittent nulling also manifests itself in the fluctuation spectra as a low-frequency feature as shown in Figs.~\ref{fig:LRFS} and \ref{fig:2DFS}.

To infer the nulling fraction (NF) of this pulsar, we adopt the Bayesian framework described by \cite{kwo+26}, modelling the observed pulse energies by incorporating an explicit NF alongside a single- or multi-component intrinsic energy distribution.
For the census observation with the greatest sensitivity, we infer an NF of $3.2 \pm 0.6$ per cent and find that the intrinsic energy distribution is best described by a two-component log-normal model (see \citealt{kwo+26} for more details and results).

We systematically applied this Bayesian fitting procedure to each individual observation, which differ in duration and in the number of antennas used (and hence in sensitivity), to infer their respective NFs.
The number of null pulses for each observation was then derived by multiplying the inferred NF by the total pulse number.
Across our 65 observations spanning seven years, the NF fluctuates from near-zero to about 55 per cent ($36.1 \pm 16.3$ nulls out of 66 pulsar rotations), as presented in \tab{tab:obs}.
Furthermore, the nulling time-scale varies considerably; while some nulls persist for only a few rotations, as shown in \fig{fig:pulsestack}, others extend over several tens of rotations.

By summing the inferred number of null pulses across all observations, we determine an overall NF of $8.1 \pm 0.5$ per cent, corresponding to $573 \pm 35$ nulls out of 7076 total pulsar rotations.
This result falls within the broad range of literature values.
For instance, \cite{bmm17} reported relatively low NFs of 2.94 per cent (at 333 MHz) and $5.6 \pm 0.7$ per cent (at 618 MHz), while \cite{njm+17} found much higher NFs of $29 \pm 4$ per cent and $30 \pm 1$ per cent at comparable observing frequencies.
\cite{gbm+26} derived the NF from a single observation (300--400 MHz) and found that it can range from $9.13 \pm 0.79$ per cent to as high as $76.41 \pm 8.75$ per cent, depending on the analysis methods used.

\section{Discussion}
\label{sec:discussion}

\subsection{Carousel-like model}
\label{sec:feedback}
The simplest form of a rotating carousel, as predicted by the \citetalias{rs75} model, is only expected to produce a single $1/P_3$ frequency (and possibly its harmonics). This is insufficient to explain the multiple co-existing periodicities observed in PSR~J1901$-$0906,
and their associated beat relations 
detected in Figs.~\ref{fig:scatter-plot-F1-F2} and \ref{fig:scatter-plot-F1-F3}.
When the beamlets of the carousel have fixed but unequal intensity, as discussed in  \cite{dr01}, co-existing spectral features can be expected. However, their $1/P_2$ and $1/P_3$ frequencies must be harmonically related \citepalias[see][]{hww+25}.
No such harmonic relation between F1, F2 and F3 is present in PSR~J1901$-$0906, and hence an alternative interpretation is required.

As pointed out in \sect{sec:beat}, the complex subpulse modulation patterns in C2 of PSR~J1901$-$0906 can be explained by a beat system in which drifting subpulses (F1) are modulated by either F2 or F3 (it is hard to mathematically distinguish).
Accordingly, we here adopt the feedback model of \cite{wri22}, as it has been shown capable of explaining the rapid amplitude modulation superimposed on the slow drifting subpulses in PSR~J1514$-$4834 \citepalias{hww+25}.
In what follows, we demonstrate that the coexistence of the multiple spectral features of PSR~J1901$-$0906, and the way they are observed to be related, can be reproduced with a single interconnected physical system.
This model is essentially the same as proposed for PSR~J1514$-$4834, with higher-order harmonics playing a significant role here.

\begin{figure}
    \centering
    \includegraphics[width=\linewidth]{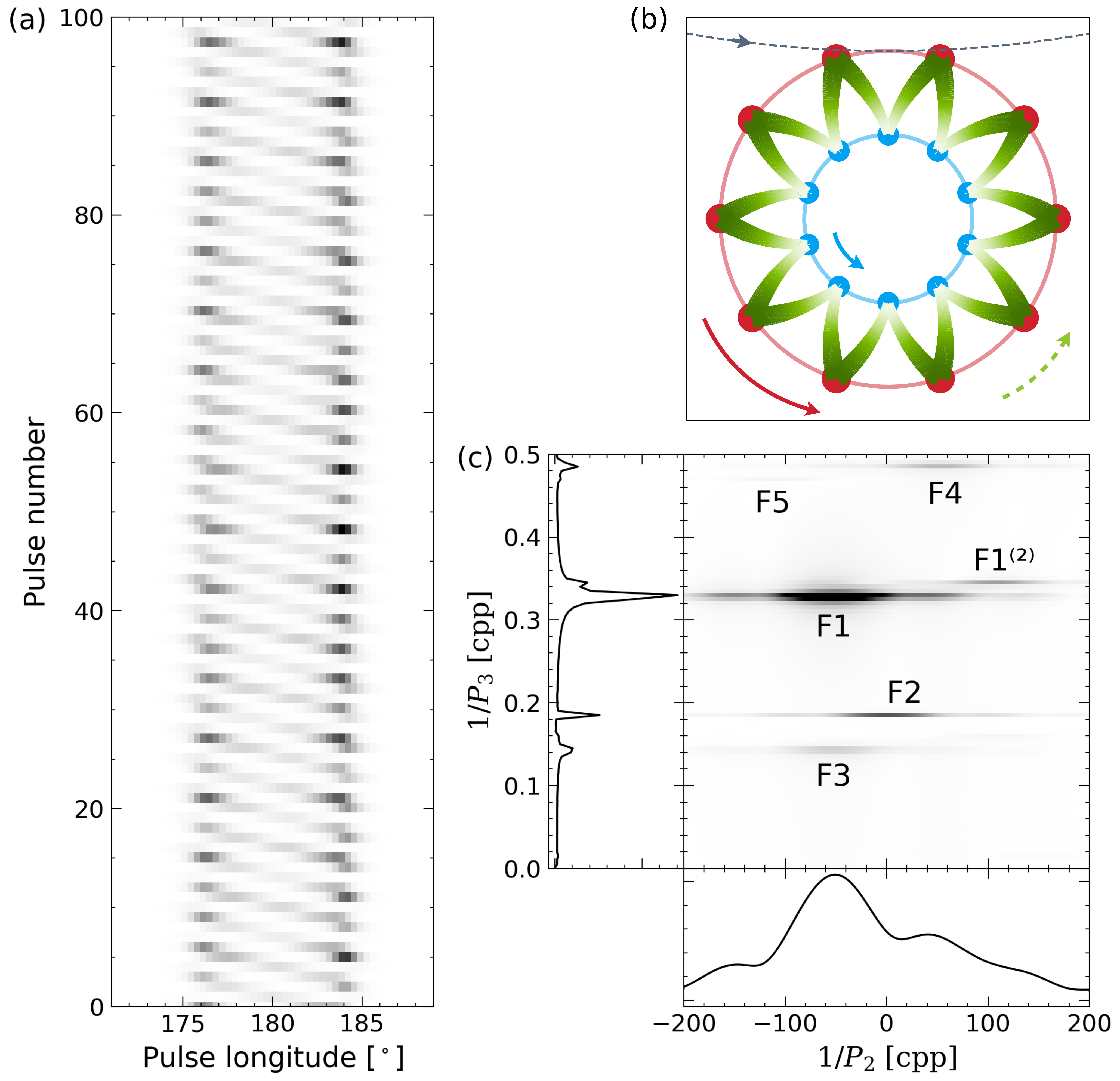} \\ [0.5em]
    \includegraphics[width=\linewidth]{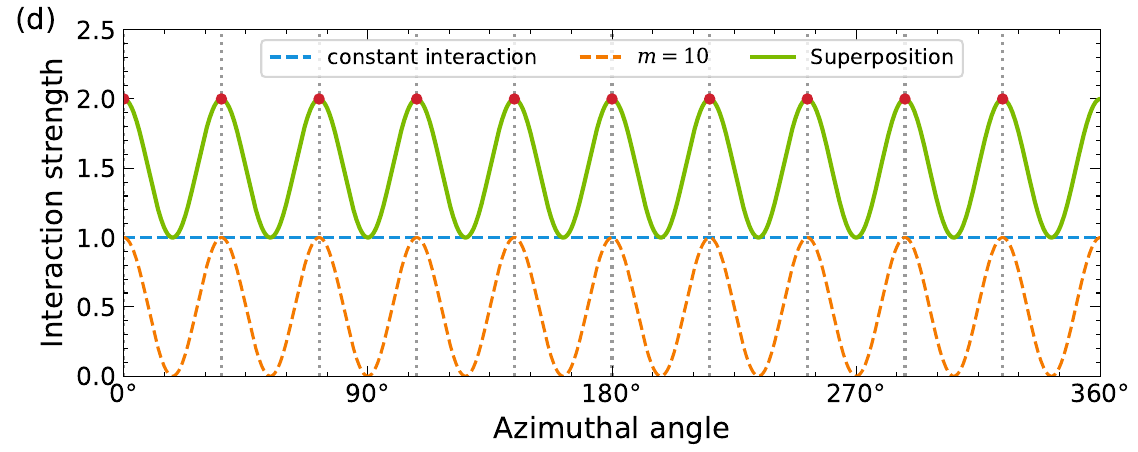}
    \caption{Simulation of the feedback model with $k=-1$, $n=10$, $P_5=70$, $\tau=5.4$. The interaction is sinusoidally modulated with $m=10$ but includes an additional constant term in the interaction (see main text for their definition). The model aims to explain properties of, and relations between, different modulation features in PSR~J1901$-$0906. It does not aim to explain differences in the modulation pattern in the two profile components.
    (a): A section of the simulated pulse stack generated using this feedback model configuration and subsequently multiplied by the observed pulse profile shape.
    (b): A geometric representation of the system. The width and darkness of the green curved line indicate the interaction strength (particle flow) between inner (blue) and outer (red) carousels.
    The arrows indicate the circulation for both the footprints (red and blue) and the particles in the flow (dashed green).
    The dashed black curve is part of the trace of the (unconstrained from observations) line of sight in the emission region.
    (c): The corresponding 2DFS for C2 of the simulated pulse stack with spectral features being labelled.
    The FFT length is 200 pulse periods.
    (d): The interaction strength as a function of the azimuthal angle for constant interaction (dashed blue, constant), sinusoidal interaction with $m=10$ (dashed orange), and their superposition (solid green), at a given moment in time. This interaction strength corresponds to the thickness of the green curve in panel (b), but unwrapped.
    The vertical dotted lines mark the azimuthal positions of the 10 footprints in the outer carousel at a particular instant.
    Thus, the intensity of the footprints at that instant is represented by the red dots on the green curve.
    }
    \label{fig:feedback}
\end{figure}

The proposed model for PSR~J1901$-$0906 is illustrated in panel (b) of \fig{fig:feedback}.
In this model, the inner (blue) and outer (red) carousels interact with each other in the form of an exchange of particle flow (green).
This particle flow, in many ways, is the only physical structure in the system that has any bearing on the observed subpulse modulation patterns. While streaming back and forth between the carousels, the particles circulate around the magnetic axis due to an $\boldsymbol{E} \boldsymbol{\times} \boldsymbol{B}$ drift, as indicated by the dashed green arrow.
The circulation period is $P_5$, and it takes a period of $\tau$ to stream back and forth between the two carousels.
Consequently, the particle flow leaves footprints on the carousels (red and blue).
These footprints resemble a carousel as in the classic \citetalias{rs75} model and are assumed to be responsible for the observed radio emission, resulting in the observed drifting subpulses.
Although the footprints circulate, unlike the \citetalias{rs75} model, their circulation is not directly determined by the $\boldsymbol{E} \boldsymbol{\times} \boldsymbol{B}$ drift.
Instead, as described by Eq.~(2) in \cite{wri22}, the observed $1/P_3$ frequency corresponding to the footprint circulation can be expressed as a beat frequency between the interaction time delay $\tau$ and the particle circulation time $P_5$ such that
\begin{equation}
    \left(\frac{1}{P_3}\right)_\mathrm{footprint}=\frac{n}{P_5}-\frac{k}{\tau}.
    \label{eq:feedback}
\end{equation}
Here, the integer $n$ represents the number of footprints in both the inner and outer carousels.
The integer $k$ indicates that a given particle in the flow (green) originating from footprint $i$ in the outer carousel (red) will return to the location of footprint $i+k$ after a time interval $\tau$, having reflected at the inner carousel (blue).
A negative $k$ indicates that the footprint circulation is faster than the particle circulation, as is the case in \fig{fig:feedback}, whereas a positive $k$ indicates the particle circulation is faster (see \apdx{apdx:k} for more explanations and examples).

A carousel with $n$ footprints, evenly spaced in azimuth and circulating with a period of $P_4$, yields drifting subpulses at $1/P_3=n/P_4$.
If the particle flow is periodic rather than uniform, i.e. if the interaction is not constant, the intensity of radiation produced by the footprints varies periodically, resulting in an additional spectral modulation feature.
Specifically, as illustrated in panels (b) and (d) in \fig{fig:feedback}, this flow modulation is assumed to depend sinusoidally\footnote{See \apdx{apdx:non-sinusoidal} for a generalization.} on the azimuthal angle, characterized by an integer $m$, such that it is $m$ times strong and $m$ times weak over one complete azimuthal cycle.
As explored in \citetalias{hww+25}, for this feature to be a pure amplitude modulation, i.e. $1/P_2=0$~cpp, the modulation of the emission produced by $n$ footprints must be synchronized: all footprints are bright or weak simultaneously.
This requires $m$ to be an exact multiple of $n$, resulting in an amplitude modulation frequency of $1/P_3=m|k|/(n\tau)$ (see \apdx{apdx:m=n}).

\fig{fig:feedback} illustrates an application of the feedback model configured to reproduce F1, F2, and F3 features, along with their various beats, as observable in C2 in PSR~J1901$-$0906. The aim is to explain and illustrate the origin of the spectral features and their relation, without aiming to explicitly model the differences in C1 and C2.
In this specific setup\footnote{$P_2$ depends on $n$ as well as things like the magnetic inclination angle. Therefore $n$, as well as $P_5$, are ill constrained from the data.},
we adopt $k=-1$, $n=10$, $\tau=5.4$, and $P_5=70$, but this is not a unique solution.
Substituting these values into \equ{eq:feedback} yields an apparent $P_3=3.05$ for the footprint circulation, corresponding to F1 and its harmonics, as shown in panel (c) in \fig{fig:feedback}.
The fact that harmonics are much weaker than the fundamental feature implies relatively broad footprints. In the Fourier domain, this finite width acts as a spatial low-pass filter, multiplying the spectrum by an attenuating envelope centred at $1/P_2=0$~cpp. As a result, spectral features with larger $|1/P_2|$ will be subjected to stronger attenuation and thus appear significantly weaker in the power 2DFS.

Apart from F1 and $\rm F1^{(2)}$ associated with the footprint circulation, by employing an interaction modulation of the particle flow with $m=n$, the amplitude modulation feature emerges at $1/P_3=m|k|/(n\tau)=0.185$~cpp and $1/P_2=0$~cpp, corresponding\footnote{Alternatively, the spectral feature associated with this interaction modulation can be identified as F3, which then modulates F1 to produce F2. A detailed demonstration of this formal equivalence is provided in \apdx{apdx:m=n}.} to F2.
This amplitude-modulates F1 and its harmonics and therefore produces beat features, as simulated in the power 2DFS shown in panel (c) of \fig{fig:feedback}, and as observed in panels (2a--2c) and (3a--3c) of \fig{fig:scatter-plot-F1-F2}.

While the modulation of the particle flow with $m=n$ is sufficient to establish the presence of all spectral features observed in PSR~J1901$-$0906 and the associated beat relations, we further include the constant interaction (see \apdx{apdx:m=0}) to accommodate the observational dominance of F1 in C2.
This is because for the constant interaction, only the drift feature F1 corresponding to the footprint circulation is observable without any additional modulation. Therefore, including the constant interaction boosts the relative strength of F1 compared to F2 and F3.
More generally, the strength ratios between these features can be tuned by introducing more complicated interaction forms or even multiple modulation particle flows, as discussed below.

Despite successfully explaining the complicated beat relations observed in C2 for PSR~J1901$-$0906, in the simulation presented in \fig{fig:feedback}, F3 and F4, the two beat features between F1 and F2, emerge as a symmetric pair with equal intrinsic amplitudes in panel (c). This is as expected from the model \citepalias[see][]{hww+25}.
However, F4 is observed to be much weaker than F3. F4 is undetectable in the observed power spectrum in \fig{fig:2DFS}, and only detectable by exploiting the Fourier phase information (see panel (2c) of \fig{fig:scatter-plot-F1-F2}).
The relative strength of F4 could be suppressed in the model by introducing a second modulation particle flow such that two sets of azimuthally offset footprints are produced on the same carousel. Although each set possesses its own internally synchronized amplitude modulation, they are not necessarily synchronized with each other.
It can be shown that for specific configurations with the same $n$ and $m$, the relative strengths of F3 and F4 can be arbitrarily tuned without disrupting the observed beat relations.
The introduction of two azimuthally offset carousels is well-motivated. In C2, we observe that the OPM switches are synchronized with the drifting subpulses (F1), see \fig{fig:p3fold}. 
This phenomenon is usually explained by the superposition of two azimuthally offset carousels, each corresponding to an OPM \citep[e.g.][]{rr03, iwj+20}.
This is further discussed in \sect{sec:pol_discussion}.

Likewise, if the model from \fig{fig:feedback} is to be applied to C1, the primary challenge to address is that F1 is observationally much weaker than F2 and F3, as seen in \fig{fig:2DFS}. 
This is likely to explain the non-detection of any beat features between F1 and the other spectral features (\fig{fig:corr-bar}).
To suppress the strength of F1 (associated with footprint circulation) while preserving the amplitude-modulation feature F2, one possible scenario is that the footprints are less regular. They might be elongated, distorted, or subjected to random localized jitter superimposed on their circulating motion. The ultimate effect of this disruption is that coherent, stable drifting subpulses (F1) can no longer form.
Crucially, even if the individual footprints become highly irregular, their collective amplitude modulation caused by the particle flows remains globally synchronized, ensuring that F2 persists as a significant spectral feature.

\subsection{Periodic polarization variability}
\label{sec:pol_discussion}

In \sect{sec:pol}, we demonstrate that the polarization state of PSR~J1901$-$0906 is modulated synchronously with the three co-existing subpulse modulations observed in total intensity.
While such synchronous modulation has been previously identified in only a small number of other pulsars (see \sect{sec:intro} and references therein), none of them exhibit co-existing periodicities.

Apart from the collinear oscillation within the $Q$-$U$ plane as expected for OPM mixing at the leading edge of C1, the polarization state at C2 traces an elliptical path in the Poincar\'e space.
The centre of this path is significantly offset from the origin and involves the periodic modulation of circular polarization $V$, which cannot be explained by incoherent OPM mixing alone.
Similar phenomena have been reported in, for example, PSRs~B0809$+$74, B0320$+$39, B0818$-$13 \citep{edw04} and B1919$+$21 \citep{ptv+22, cas23}.

To explain the periodic switching between OPMs synchronized with the drifting subpulses within the framework of the \citetalias{rs75} model, \cite{rr03} suggest the emission from the underlying rotating carousel is split into two `images', each corresponding to one of the observed OPMs. Such a mechanism could equally well work with the feedback model discussed in \sect{sec:feedback}.
Building upon the concept of producing split images, \cite{ptv+22} proposed a four-mode mixing model to explain the periodically modulated elliptical trajectory of the polarization state in Poincar\'e space observed in PSR~B1919$+$21. In this interpretation, two similar carousels are required, each generating a pair of OPMs. These two sets of OPMs are separated by nearly $90^\circ$ in Poincar\'e space, such that the observed emission results from the incoherent superposition of these four modes.
Although we assume that the inner carousel in PSR~J1901$-$0906 is not directly observable under our feedback model for drifting subpulses (\sect{sec:feedback}), the necessity of having two carousels is compatible with the four-mode mixing model.

\section{Conclusion}
\label{sec:conclusion}

In this paper, we carried out a comprehensive analysis of the complex subpulse modulation patterns in PSR~J1901$-$0906 using $L$-band data from the TPA programme. We identified the coexistence of seven subpulse modulation periodicities, of which the most prominent are labelled F1, F2, and F3 and are detectable in both profile components.
These periodicities are not independent but obey a beat relation, $\rm F1 = F2 + F3$, in terms of their $1/P_2$ and $1/P_3$ frequencies (\sect{sec:F123}), as well as their subpulse phases (\sect{sec:subpulse_phase}).
These co-existing periodicities modulate each other in the form of a beat system, similar to PSR~J1514$-$4834.
As a consequence of the modulation, additional beat features are expected and detected by applying the Fourier phase correlation technique developed in \citetalias{hww+25} (\sect{sec:beat}).
This analysis establishes that these spectral features originate from an underlying, interconnected physical system rather than independent processes. 

Furthermore, we applied a $P_3$-folding technique to demonstrate that the single-pulse polarization states of this pulsar are modulated and synchronized with F1, F2, and F3 (\sect{sec:pol}).
Reported synchronized polarization modulation is relatively rare in the population.
Notably, PSR~J1901$-$0906 is the only pulsar among these involving multiple co-existing periodicities.
In addition to these complex subpulse modulations, the pulsar exhibits intermittent nulling with an overall nulling fraction of $8.1 \pm 0.5$ per cent (\sect{sec:null}).

Our observations of PSR~J1901$-$0906 reveal a much more complex subpulse modulation that cannot be explained by the widely invoked carousel model in its original form. Alternatively, we demonstrate that the feedback model \citep{wri22} is capable of generating F1, F2, and F3 observed in this pulsar, with their beat relations, as well as their resulting beat features, emerging naturally (\sect{sec:feedback}). 
Just as this model has been successfully applied to explain the co-existing periodicities in PSR~J1514$-$4834 \citepalias{hww+25}, it can equally be applied to PSR~J1901$-$0906. The main difference is that in the case of PSR~J1901$-$0906 higher-order harmonics play a vital role within the same physical framework.

\section*{Acknowledgements}
The MeerKAT telescope is operated by the South African Radio Astronomy Observatory, which is a facility of the National Research Foundation, an agency of the Department of Science and Innovation.
Pulsar research at Jodrell Bank Centre for Astrophysics and Jodrell Bank Observatory is supported by a consolidated grant from the UK Science and Technology Facilities Council (STFC). 
JAH acknowledges funding from the STFC Doctoral Training studentship.
GW thanks the University of Manchester for Visitor status. 
XS acknowledges support through project CORTEX (NWA.1160.18.316) of the research programme NWA-ORC which is financed by the Dutch Research Council (NWO).
MeerTime data are housed and processed on the OzSTAR supercomputer at Swinburne University of Technology.
We thank the anonymous reviewer for their valuable comments and suggestions, which helped improve the clarity and quality of this manuscript.

\section*{Data Availability}
Data underlying this article will be shared upon reasonable request to the corresponding author.


\bibliographystyle{mnras}
\bibliography{references}

@article{hww+25,
  title = {The Thousand-Pulsar-Array programme on MeerKAT XVII. Discovery of beating radio emission variability in PSR J1514-4834},
  shorttitle = {The Thousand-Pulsar-Array programme on MeerKAT XVII},
  author = {Hsu, J A and Weltevrede, P and Wright, G and Keith, M J and Oswald, L S and Song, X and Wang, H},
  year = {2025},
  month = feb,
  journal = {\mnras},
  volume = {538},
  number = {2},
  pages = {1063--1080},
  doi = {10.1093/mnras/staf323}
}

@article{dr01,
  title = {The topology and polarization of sub-beams associated with the 'drifting' sub-pulse emission of pulsar B0943+10-I. Analysis of Arecibo 430- and 111-MHz observations},
  author = {Deshpande, A. A. and Rankin, J. M.},
  year = {2001},
  month = apr,
  journal = {\mnras},
  volume = {322},
  number = {3},
  pages = {438--460},
  doi = {10.1046/j.1365-8711.2001.04079.x}
}

@article{njm+17,
  title = {Simultaneous multi-frequency single pulse observations of pulsars},
  author = {Naidu, A. and Joshi, B. C. and Manoharan, P. K. and KrishnaKumar, M. A.},
  year = {2017},
  month = aug,
  journal = {\aap},
  volume = {604},
  pages = {A45},
  doi = {10.1051/0004-6361/201629937}
}

@article{bmm+19,
  title = {Classification of subpulse drifting in pulsars},
  author = {Basu, Rahul and Mitra, Dipanjan and Melikidze, George I and Skrzypczak, Anna},
  year = {2019},
  month = jan,
  journal = {\mnras},
  volume = {482},
  number = {3},
  pages = {3757--3788},
  doi = {10.1093/mnras/sty2846}
}

@article{sws+23,
  title = {The Thousand-Pulsar-Array programme on MeerKAT - VIII. The subpulse modulation of 1198 pulsars},
  author = {Song, X. and Weltevrede, P. and Szary, A. and Wright, G. and Keith, M. J. and Basu, A. and Johnston, S. and Karastergiou, A. and Main, R. A. and Oswald, L. S. and Parthasarathy, A. and Posselt, B. and Bailes, M. and Buchner, S. and Hugo, B. and Serylak, M.},
  year = {2023},
  month = apr,
  journal = {\mnras},
  volume = {520},
  pages = {4562--4581},
  doi = {10.1093/mnras/stad135}
}

@article{wes06,
  title = {The subpulse modulation properties of pulsars at 21 cm},
  author = {Weltevrede, P. and Edwards, R. T. and Stappers, B. W.},
  year = 2006,
  month = jan,
  journal = {\aap},
  volume = {445},
  number = {1},
  pages = {243--272},
  doi = {10.1051/0004-6361:20053088}
}

@article{wse07,
  title = {The subpulse modulation properties of pulsars at 92 cm and the frequency dependence of subpulse modulation},
  author = {Weltevrede, P. and Stappers, B. W. and Edwards, R. T.},
  year = {2007},
  month = jul,
  journal = {\aap},
  volume = {469},
  number = {2},
  pages = {607--631},
  publisher = {EDP Sciences},
  doi = {10.1051/0004-6361:20066855}
}

@article{rs75,
  title = {Theory of pulsars: polar gaps, sparks, and coherent microwave radiation.},
  shorttitle = {Theory of pulsars},
  author = {Ruderman, M. A. and Sutherland, P. G.},
  year = {1975},
  month = feb,
  journal = {\apj},
  volume = {196},
  pages = {51--72},
  doi = {10.1086/153393}
}

@article{jkk+20,
  title = {The Thousand-Pulsar-Array programme on MeerKAT -- I. Science objectives and first results},
  author = {Johnston, Simon and Karastergiou, A and Keith, M J and Song, X and Weltevrede, P and Abbate, F and Bailes, M and Buchner, S and Camilo, F and Geyer, M and Hugo, B and Jameson, A and Kramer, M and Parthasarathy, A and Reardon, D J and Ridolfi, A and Serylak, M and Shannon, R M and Spiewak, R and {van~Straten}, W and Venkatraman~Krishnan, V and Jankowski, F and Meyers, B W and Oswald, L and Posselt, B and Sobey, C and Szary, A and {van~Leeuwen}, J},
  year = {2020},
  month = apr,
  journal = {\mnras},
  volume = {493},
  number = {3},
  pages = {3608--3615},
  doi = {10.1093/mnras/staa516}
}

@article{bja+20,
  title = {The MeerKAT telescope as a pulsar facility: System verification and early science results from MeerTime},
  shorttitle = {The MeerKAT telescope as a pulsar facility},
  author = {Bailes, M. and Jameson, A. and Abbate, F. and Barr, E. D. and Bhat, N. D. R. and Bondonneau, L. and Burgay, M. and Buchner, S. J. and Camilo, F. and Champion, D. J. and Cognard, I. and Demorest, P. B. and Freire, P. C. C. and Gautam, T. and Geyer, M. and Griessmeier, J.-M. and Guillemot, L. and Hu, H. and Jankowski, F. and Johnston, S. and Karastergiou, A. and Karuppusamy, R. and Kaur, D. and Keith, M. J. and Kramer, M. and {van Leeuwen}, J. and Lower, M. E. and Maan, Y. and McLaughlin, M. A. and Meyers, B. W. and Os{\l}owski, S. and Oswald, L. S. and Parthasarathy, A. and Pennucci, T. and Posselt, B. and Possenti, A. and Ransom, S. M. and Reardon, D. J. and Ridolfi, A. and Schollar, C. T. G. and Serylak, M. and Shaifullah, G. and Shamohammadi, M. and Shannon, R. M. and Sobey, C. and Song, X. and Spiewak, R. and Stairs, I. H. and Stappers, B. W. and {van Straten}, W. and Szary, A. and Theureau, G. and Venkatraman Krishnan, V. and Weltevrede, P. and Wex, N. and Abbott, T. D. and Adams, G. B. and Burger, J. P. and Gamatham, R. R. G. and Gouws, M. and Horn, D. M. and Hugo, B. and Joubert, A. F. and Manley, J. R. and McAlpine, K. and Passmoor, S. S. and {Peens-Hough}, A. and Ramudzuli, Z. R and Rust, A. and Salie, S. and Schwardt, L. C. and Siebrits, R. and Van Tonder, G. and Van Tonder, V. and Welz, M. G.},
  year = {2020},
  month = jul,
  journal = {\pasa},
  volume = {37},
  pages = {e028},
  doi = {10.1017/pasa.2020.19}
}

@article{DSPSR,
  title = {DSPSR: Digital Signal Processing Software for Pulsar Astronomy},
  shorttitle = {DSPSR},
  author = {{van Straten}, W. and Bailes, M.},
  year = {2011},
  journal = {\pasa},
  volume = {28},
  number = {1},
  pages = {1--14},
  doi = {10.1071/AS10021}
}

@article{es02,
  title = {Drifting sub-pulse analysis using the two-dimensional Fourier transform},
  author = {Edwards, R. T. and Stappers, B. W.},
  year = {2002},
  month = oct,
  journal = {\aap},
  volume = {393},
  number = {2},
  pages = {733--748},
  publisher = {EDP Sciences},
  doi = {10.1051/0004-6361:20021067}
}

@article{PSRSALSA,
  title = {Investigation of the bi-drifting subpulses of radio pulsar B1839-04 utilising the open-source data-analysis project PSRSALSA},
  author = {Weltevrede, P.},
  year = {2016},
  month = may,
  journal = {\aap},
  volume = {590},
  pages = {A109},
  doi = {10.1051/0004-6361/201527950}
}

@article{ssw09,
  title = {S2DFS: analysis of temporal changes of drifting subpulses},
  shorttitle = {S2DFS},
  author = {Serylak, M. and Stappers, B. W. and Weltevrede, P.},
  year = {2009},
  month = nov,
  journal = {\aap},
  volume = {506},
  number = {2},
  pages = {865--874},
  publisher = {EDP Sciences},
  doi = {10.1051/0004-6361/200912453}
}

@article{LRFS,
  title = {Correlation of Subpulse Structure in a Sequence of Pulses from Pulsar PSR 1919+21},
  author = {Backer, D. C.},
  year = {1970},
  month = aug,
  journal = {Nature},
  volume = {227},
  pages = {692--695},
  doi = {10.1038/227692a0}
}

@article{wkw+25,
  title = {The Thousand-Pulsar-Array programme on MeerKAT -- XVIII. Complex sub-pulse modulation patterns, bi-drifting and mode changing of nine radio pulsars},
  author = {Wang, H and Keith, M J and Weltevrede, P and Wright, G and Bhattacharyya, B and Hsu, J A and Song, X},
  year = 2025,
  month = oct,
  journal = {\mnras},
  volume = {544},
  number = {1},
  pages = {234--254},
  doi = {10.1093/mnras/staf1710}
}

@article{smk05,
  title = {Frequency dependence of the drifting subpulses of PSR B0031-07},
  author = {Smits, J. M. and Mitra, D. and Kuijpers, J.},
  year = 2005,
  month = sep,
  journal = {\aap},
  volume = {440},
  pages = {683--692},
  doi = {10.1051/0004-6361:20041626}
}

@article{es03a,
  title = {The frequency-dependence of drifting subpulse patterns},
  author = {Edwards, R. T. and Stappers, B. W.},
  year = 2003,
  month = nov,
  journal = {\aap},
  volume = {410},
  number = {3},
  pages = {961--966},
  publisher = {EDP Sciences},
  doi = {10.1051/0004-6361:20031326}
}

@article{iwj+20,
  title = {The drifting sub-pulses of PSR B0031-07 and its synchronously modulated radio polarization},
  author = {Ilie, C. D. and Weltevrede, P. and Johnston, S. and Chen, T.},
  year = 2020,
  month = jan,
  journal = {\mnras},
  volume = {491},
  pages = {3385--3394},
  publisher = {OUP},
  doi = {10.1093/mnras/stz3167}
}

@article{es04,
    title = {Ellipticity and deviations from orthogonality in the polarization modes of {PSR} {B0329}+54},
    volume = {421},
    copyright = {© ESO, 2004},
    issn = {0004-6361, 1432-0746},
    url = {https://www.aanda.org/articles/aa/abs/2004/26/aa0228-04/aa0228-04.html},
    doi = {10.1051/0004-6361:20040228},
    language = {en},
    number = {2},
    urldate = {2025-04-01},
    journal = {\aap},
    author = {Edwards, R. T. and Stappers, B. W.},
    month = jul,
    year = {2004},
    note = {Number: 2
Publisher: EDP Sciences},
    pages = {681--691},
}

@article{edw04,
  title = {The polarization of drifting subpulses},
  author = {Edwards, R. T.},
  year = 2004,
  month = nov,
  journal = {\aap},
  volume = {426},
  number = {2},
  pages = {677--686},
  doi = {10.1051/0004-6361:20041029}
}

@phdthesis{cas23,
  title = {A Spectral Analysis of the Polarisation of Drifting Sub-Pulses},
  author = {Castelltort Schnaas, Ander},
  year = 2023,
  school = {Auckland University of Technology},
  url = {http://hdl.handle.net/10292/17517}
}

@article{ptv+22,
  title = {The polarisation of the drifting sub-pulses from PSR B1919+21},
  author = {Primak, N. and Tiburzi, C. and {van Straten}, W. and Dyks, J. and Gulyaev, S.},
  year = 2022,
  month = jan,
  journal = {\aap},
  volume = {657},
  pages = {A34},
  doi = {10.1051/0004-6361/202140811}
}

@article{rr03,
  title = {Toward an Empirical Theory of Pulsar Emission. VIII. Subbeam Circulation and the Polarization-Modal Structure of Conal Beams},
  author = {Rankin, Joanna M. and Ramachandran, R.},
  year = 2003,
  month = jun,
  journal = {\apj},
  volume = {590},
  number = {1},
  pages = {411--423},
  doi = {10.1086/374884}
}

@article{rrs+02,
  title = {Pulsar ``drifting''-subpulse polarization: No evidence for systematic polarization-angle rotations},
  shorttitle = {Pulsar ``drifting''-subpulse polarization},
  author = {Ramachandran, R. and Rankin, J. M. and Stappers, B. W. and Kouwenhoven, M. L. A. and Van Leeuwen, A. G. J.},
  year = 2002,
  month = jan,
  journal = {\aap},
  volume = {381},
  number = {3},
  pages = {993--999},
  doi = {10.1051/0004-6361:20011525}
}

@article{thh+71,
       author = {{Taylor}, J.~H. and {Huguenin}, G.~R. and {Hirsch}, R.~M. and {Manchester}, R.~N.},
        title = "{Polarization of the Drifting Subpulses of Pulsar 0809 + 74}",
      journal = {\aplett},
         year = 1971,
        month = aug,
       volume = {9},
        pages = {205-208},
       adsurl = {https://ui.adsabs.harvard.edu/abs/1971ApL.....9..205T}
}

@article{ew01,
  title = {Emission Beam Geometry of Selected Pulsars Derived from Average Pulse Polarization Data},
  author = {Everett, J. E. and Weisberg, J. M.},
  year = 2001,
  month = may,
  journal = {\apj},
  volume = {553},
  number = {1},
  pages = {341--357},
  doi = {10.1086/320652}
}

@article{wk74,
  title = {The linear polarization of quasi-stellar radio sources at 3.71 and 11.1 centimeters},
  author = {Wardle, J. F. C. and Kronberg, P. P.},
  year = 1974,
  month = dec,
  journal = {\apj},
  volume = {194},
  pages = {249},
  doi = {10.1086/153240}
}

@article{mth75,
  title = {Observations of pulsar radio emission. II - Polarization of individual pulses},
  author = {Manchester, R. N. and Taylor, J. H. and Huguenin, G. R.},
  year = 1975,
  month = feb,
  journal = {\apj},
  volume = {196},
  pages = {83},
  doi = {10.1086/153395}
}

@article{brc76,
  title = {Orthogonal mode emission in geometric models of pulsar polarisation},
  author = {Backer, D. C. and Rankin, John M. and Campbell, D. B.},
  year = 1976,
  month = sep,
  journal = {Nature},
  volume = {263},
  number = {5574},
  pages = {202--207},
  doi = {10.1038/263202a0}
}

@article{dwi21,
  title = {Circular polarization in radio pulsar PSR B1451-68: coherent mode transitions and intrabeam interference},
  shorttitle = {Circular polarization in radio pulsar PSR B1451-68},
  author = {Dyks, J and Weltevrede, P and Ilie, C},
  year = 2021,
  month = jan,
  journal = {\mnras},
  volume = {501},
  number = {2},
  pages = {2156--2173},
  doi = {10.1093/mnras/staa3762}
}

@article{swv+26,
  title = {Main and inter-pulse interaction in PSRs J1842+0358 and J1926+0737: evidence of inter-pole communication},
  shorttitle = {Main and inter-pulse interaction in PSRs J1842+0358 and J1926+0737},
  author = {Song, X. and Weltevrede, P. and Van Leeuwen, J. and Wright, G. and Keith, M.},
  year = 2026,
  month = feb,
  journal = {\aap},
  volume = {706},
  pages = {A160},
  doi = {10.1051/0004-6361/202555190}
}

@article{dc68,
  title = {Second Periodic Pulsation in Pulsars},
  author = {Drake, F. D. and Craft, H. D.},
  year = 1968,
  month = oct,
  journal = {Nature},
  volume = {220},
  number = {5164},
  pages = {231--235},
  publisher = {Nature Publishing Group},
  doi = {10.1038/220231a0}
}

@article{em69,
  title = {Polarization of Pulsating Radio Sources},
  author = {Ekers, R. D. and Moffet, A. T.},
  year = 1969,
  month = oct,
  journal = {\apj},
  volume = {158},
  pages = {L1},
  doi = {10.1086/180419}
}

@article{ms77,
  title = {Generalised Kramers-Kronig formula for spatially dispersive media},
  author = {Melrose, D B and Stoneham, R J},
  year = 1977,
  month = jan,
  journal = {Journal of Physics A: Mathematical and General},
  volume = {10},
  number = {1},
  pages = {L17-L20},
  doi = {10.1088/0305-4470/10/1/004}
}

@article{cp76,
  title = {Theory of the polarization of pulsar radio radiation},
  author = {Cocke, W. J. and Pacholczyk, A. G.},
  year = 1976,
  month = feb,
  journal = {\apj},
  volume = {204},
  pages = {L13},
  doi = {10.1086/182044}
}

@article{ab86,
  title = {Wave propagation in pulsar magnetospheres - Dispersion relations and normal modes of plasmas in superstrong magnetic fields},
  author = {Arons, J. and Barnard, J. J.},
  year = 1986,
  month = mar,
  journal = {\apj},
  volume = {302},
  pages = {120},
  doi = {10.1086/163978}
}

@article{ran86,
  title = {Toward an Empirical Theory of Pulsar Emission. III. Mode Changing, Drifting Subpulses, and Pulse Nulling},
  author = {Rankin, J. M.},
  year = 1986,
  month = feb,
  journal = {\apj},
  volume = {301},
  pages = {901},
  doi = {10.1086/163955}
}

@article{swk+21,
  title = {The Thousand-Pulsar-Array programme on MeerKAT - II. Observing strategy for pulsar monitoring with subarrays},
  author = {Song, X. and Weltevrede, P. and Keith, M. J. and Johnston, S. and Karastergiou, A. and Bailes, M. and Barr, E. D. and Buchner, S. and Geyer, M. and Hugo, B. V. and Jameson, A. and Parthasarathy, A. and Reardon, D. J. and Serylak, M. and Shannon, R. M. and Spiewak, R. and {van Straten}, W. and Venkatraman Krishnan, V.},
  year = 2021,
  month = aug,
  journal = {\mnras},
  volume = {505},
  pages = {4456--4467},
  doi = {10.1093/mnras/staa3805}
}

@article{lml+98,
  title = {The Parkes Southern Pulsar Survey --- II. Final results and population analysis},
  author = {Lyne, A. G. and Manchester, R. N. and Lorimer, D. R. and Bailes, M. and D'Amico, N. and Tauris, T. M. and Johnston, S. and Bell, J. F. and Nicastro, L.},
  year = 1998,
  month = apr,
  journal = {\mnras},
  volume = {295},
  number = {4},
  pages = {743--755},
  doi = {10.1046/j.1365-8711.1998.01144.x}
}

@article{kjk+24,
  title = {The Thousand-Pulsar-Array programme on MeerKAT - XIII. Timing, flux density, rotation measure, and dispersion measure time series of 597 pulsars},
  author = {Keith, M. J. and Johnston, S. and Karastergiou, A. and Weltevrede, P. and Lower, M. E. and Basu, A. and Posselt, B. and Oswald, L. S. and Parthasarathy, A. and Cameron, A. D. and Serylak, M. and Buchner, S.},
  year = 2024,
  month = may,
  journal = {\mnras},
  volume = {530},
  pages = {1581--1591},
  publisher = {OUP},
  doi = {10.1093/mnras/stae937}
}

@article{zhm00,
  title = {Radio Pulsar Death Line Revisited: Is PSR J2144-3933 Anomalous?},
  shorttitle = {Radio Pulsar Death Line Revisited},
  author = {Zhang, Bing and Harding, Alice K. and Muslimov, Alexander G.},
  year = 2000,
  month = mar,
  journal = {\apj},
  volume = {531},
  pages = {L135-L138},
  doi = {10.1086/312542}
}

@article{wri22,
  title = {Pulsar emission patterns seen as evidence for magnetospheric interactions},
  author = {Wright, Geoff},
  year = 2022,
  month = jun,
  journal = {\mnras},
  volume = {514},
  number = {3},
  pages = {4046--4060},
  doi = {10.1093/mnras/stac1629}
}

@article{hrh+26,
  title = {A binary model of long-period radio transients and white dwarf pulsars},
  author = {Horv{\'a}th, Csan{\'a}d and Rea, Nanda and {Hurley-Walker}, Natasha and McSweeney, Samuel J. and Perley, Richard A. and Lenc, Emil},
  year = 2026,
  month = apr,
  journal = {Nature Astronomy},
  volume = {10},
  number = {4},
  pages = {522--530},
  publisher = {Nature Publishing Group},
  doi = {10.1038/s41550-025-02760-y}
}

@article{reh+25,
  title = {A Link Between White Dwarf Pulsars and Polars: Multiwavelength Observations of the 9.36-minute Period Variable Gaia22ayj},
  shorttitle = {A Link Between White Dwarf Pulsars and Polars},
  author = {Rodriguez, Antonio C. and {El-Badry}, Kareem and Hakala, Pasi and {Rodr{\'i}guez-Gil}, Pablo and Bao, Tong and Galiullin, Ilkham and Kurlander, Jacob A. and Law, Casey J. and Pelisoli, Ingrid and Schreiber, Matthias R. and Burdge, Kevin and Caiazzo, Ilaria and Roestel, Jan Van and Szkody, Paula and Drake, Andrew J. and Buckley, David A. H. and Potter, Stephen B. and Gaensicke, Boris and Mori, Kaya and Bellm, Eric C. and Kulkarni, Shrinivas R. and Prince, Thomas A. and Graham, Matthew and Kasliwal, Mansi M. and Rose, Sam and Sharma, Yashvi and Ahumada, Tom{\'a}s and Anand, Shreya and Viitanen, Akke and Wold, Avery and Chen, Tracy X. and Riddle, Reed and Smith, Roger},
  year = 2025,
  month = feb,
  journal = {\pasp},
  volume = {137},
  number = {2},
  pages = {024202},
  doi = {10.1088/1538-3873/adb0f1}
}

@article{as85,
  title = {Is GX5 -- 1 a millisecond pulsar?},
  author = {Alpar, M. Ali and Shaham, Jacob},
  year = 1985,
  month = jul,
  journal = {Nature},
  volume = {316},
  number = {6025},
  pages = {239--241},
  doi = {10.1038/316239a0}
}

@article{mue26,
  title = {Evidence for 1.01~s pulsations of the Central Compact Object in the supernova remnant RCW 103 with ASCA, XMM-Newton, and NuSTAR},
  author = {Makishima, Kazuo and Uchida, Nagomi and Enoto, Teruaki},
  year = 2026,
  month = feb,
  journal = {\pasj},
  volume = {78},
  number = {2},
  pages = {688--709},
  doi = {10.1093/pasj/psag011}
}

@article{ntz+13,
  title = {Steps towards a solution of the FS Aurigae puzzle -- II. Confirmation of the intermediate polar status},
  author = {Neustroev, V. V. and Tovmassian, G. H. and Zharikov, S. V. and Sjoberg, George},
  year = 2013,
  month = jul,
  journal = {\mnras},
  volume = {432},
  number = {3},
  pages = {2596--2614},
  doi = {10.1093/mnras/stt622}
}

@article{war86,
  title = {Multiple optical orbital sidebands in intermediate polars},
  author = {Warner, Brian},
  year = 1986,
  month = mar,
  journal = {\mnras},
  volume = {219},
  number = {2},
  pages = {347--356},
  doi = {10.1093/mnras/219.2.347}
}

@article{tzn07,
  title = {Evidence of Precession of the White Dwarf in Cataclysmic Variables},
  author = {Tovmassian, Gaghik H. and Zharikov, Sergey V. and Neustroev, Vitaly V.},
  year = 2007,
  month = jan,
  journal = {\apj},
  volume = {655},
  number = {1},
  pages = {466--472},
  doi = {10.1086/509864}
}

@article{rwr05,
  title = {Pulsar PSR B2303+30: a single system of drifting subpulses, moding and nulling},
  shorttitle = {Pulsar PSR B2303+30},
  author = {Redman, Stephen L. and Wright, Geoffrey A. E. and Rankin, Joanna M.},
  year = 2005,
  month = mar,
  journal = {\mnras},
  volume = {357},
  number = {3},
  pages = {859--872},
  doi = {10.1111/j.1365-2966.2005.08672.x}
}

@article{htt70,
  title = {The Radio Emission from Pulsar MP 0031-07},
  author = {Huguenin, G. R. and Taylor, J. H. and Troland, T. H.},
  year = 1970,
  month = dec,
  journal = {\apj},
  volume = {162},
  pages = {727},
  doi = {10.1086/150704}
}

@article{vjv+85,
  title = {Intensity-dependent quasi-periodic oscillations in the X-ray flux of GX5-1},
  author = {Van Der Klis, M. and Jansen, F. and Van Paradijs, J. and Lewin, W. H. G and Van Den Heuvel, E. P. J. and Tr{\"u}mper, J. E. and Sztajno, M.},
  year = 1985,
  month = jul,
  journal = {Nature},
  volume = {316},
  number = {6025},
  pages = {225--230},
  doi = {10.1038/316225a0}
}

@article{szs+96,
  title = {Millisecond X-Ray Variability from an Accreting Neutron Star System},
  author = {Strohmayer, Tod E. and Zhang, William and Swank, Jean H. and Smale, Alan and Titarchuk, Lev and Day, Charles and Lee, Umin},
  year = 1996,
  month = sep,
  journal = {\apj},
  volume = {469},
  number = {1},
  pages = {L9},
  publisher = {IOP Publishing},
  doi = {10.1086/310261}
}

@article{mlp98,
  title = {Sonic-Point Model of Kilohertz Quasi-periodic Brightness Oscillations in Low-Mass X-Ray Binaries},
  author = {Miller, M. Coleman and Lamb, Frederick K. and Psaltis, Dimitrios},
  year = 1998,
  month = dec,
  journal = {\apj},
  volume = {508},
  number = {2},
  pages = {791},
  publisher = {IOP Publishing},
  doi = {10.1086/306408}
}

@article{mgh+16,
  title = {A radio-pulsing white dwarf binary star},
  author = {Marsh, T. R. and G{\"a}nsicke, B. T. and H{\"u}mmerich, S. and Hambsch, F.-J. and Bernhard, K. and Lloyd, C. and Breedt, E. and Stanway, E. R. and Steeghs, D. T. and Parsons, S. G. and Toloza, O. and Schreiber, M. R. and Jonker, P. G. and {van Roestel}, J. and Kupfer, T. and Pala, A. F. and Dhillon, V. S. and Hardy, L. K. and Littlefair, S. P. and Aungwerojwit, A. and Arjyotha, S. and Koester, D. and Bochinski, J. J. and Haswell, C. A. and Frank, P. and Wheatley, P. J.},
  year = 2016,
  month = sep,
  journal = {Nature},
  volume = {537},
  number = {7620},
  pages = {374--377},
  publisher = {Nature Publishing Group},
  doi = {10.1038/nature18620}
}

@article{md81,
  title = {Pulsar disk systems},
  author = {Michel, F. C. and Dessler, A. J.},
  year = 1981,
  month = dec,
  journal = {\apj},
  volume = {251},
  pages = {654},
  doi = {10.1086/159511}
}

@article{cs08,
  title = {Rocking the Lighthouse: Circumpulsar Asteroids and Radio Intermittency},
  shorttitle = {Rocking the Lighthouse},
  author = {Cordes, J. M. and Shannon, R. M.},
  year = 2008,
  month = aug,
  journal = {\apj},
  volume = {682},
  number = {2},
  pages = {1152--1165},
  publisher = {American Astronomical Society},
  doi = {10.1086/589425}
}

@article{wri79a,
  title = {The possibility of accretion onto radio pulsars},
  author = {Wright, G. A. E.},
  year = 1979,
  month = jul,
  journal = {Nature},
  volume = {280},
  number = {5717},
  pages = {40--41},
  publisher = {Nature Publishing Group},
  doi = {10.1038/280040a0}
}

@article{che85,
  title = {Interstellar grains and current flow in pulsar magnetospheres},
  author = {Cheng, A. F.},
  year = 1985,
  month = dec,
  journal = {\apj},
  volume = {299},
  pages = {917},
  doi = {10.1086/163758}
}

@article{lm07,
  title = {Pulsar radiation belts and transient radio emission},
  author = {Luo, Qinghuan and Melrose, Don},
  year = 2007,
  month = jul,
  journal = {\mnras},
  volume = {378},
  number = {4},
  pages = {1481--1490},
  doi = {10.1111/j.1365-2966.2007.11889.x}
}

@article{whi88,
  title = {Numerical simulations of accretion discs -- I. Superhumps: a tidal phenomenon of accretion discs},
  shorttitle = {Numerical simulations of accretion discs -- I. Superhumps},
  author = {Whitehurst, Robert},
  year = 1988,
  month = may,
  journal = {\mnras},
  volume = {232},
  number = {1},
  pages = {35--51},
  doi = {10.1093/mnras/232.1.35}
}

@article{okj23,
  title = {Pulsar polarization: a partial-coherence model},
  shorttitle = {Pulsar polarization},
  author = {Oswald, L S and Karastergiou, A and Johnston, S},
  year = 2023,
  month = aug,
  journal = {\mnras},
  volume = {525},
  number = {1},
  pages = {840--853},
  doi = {10.1093/mnras/stad2271}
}

@article{kj26,
  title = {Radio emission from beyond the light cylinder in millisecond pulsars},
  author = {Kramer, Michael and Johnston, Simon},
  year = 2026,
  month = mar,
  journal = {\mnras},
  volume = {547},
  number = {4},
  pages = {staf2258},
  doi = {10.1093/mnras/staf2258}
}

@article{dhr04,
  title = {Relativistic Effects and Polarization in Three High-Energy Pulsar Models},
  author = {Dyks, J. and Harding, Alice K. and Rudak, B.},
  year = 2004,
  month = may,
  journal = {\apj},
  volume = {606},
  number = {2},
  pages = {1125--1142},
  doi = {10.1086/383121}
}

@article{bpd+15,
  title = {A broadband radio study of the average profile and giant pulses from PSR B1821-24A},
  author = {Bilous, A. V. and Pennucci, T. T. and Demorest, P. and Ransom, S. M.},
  year = 2015,
  month = apr,
  journal = {\apj},
  volume = {803},
  number = {2},
  pages = {83},
  doi = {10.1088/0004-637X/803/2/83}
}

@article{bmm17,
  title = {Meterwavelength Single-pulse Polarimetric Emission Survey. III. The Phenomenon of Nulling in Pulsars},
  author = {Basu, Rahul and Mitra, Dipanjan and Melikidze, George I.},
  year = 2017,
  month = sep,
  journal = {\apj},
  volume = {846},
  number = {2},
  pages = {109},
  doi = {10.3847/1538-4357/aa862d}
}

@article{gbm+26,
  title = {The ONs and OFFs of Pulsar Radio Emission: Characterizing the Nulling Phenomenon},
  shorttitle = {The ONs and OFFs of Pulsar Radio Emission},
  author = {Grover, Garvit and Bhat, N. D. Ramesh and McSweeney, Samuel J. and Lee, Christopher P. and Tan, Chia Min and Fu, Shih Ching and Meyers, Bradley W.},
  year = 2026,
  month = feb,
  journal = {\apj},
  volume = {997},
  number = {2},
  pages = {164},
  doi = {10.3847/1538-4357/ae2bd9}
}

@article{bmm+16,
  title = {METERWAVELENGTH SINGLE-PULSE POLARIMETRIC EMISSION SURVEY. II. THE PHENOMENON OF DRIFTING SUBPULSES},
  author = {Basu, Rahul and Mitra, Dipanjan and Melikidze, George I. and Maciesiak, Krzysztof and Skrzypczak, Anna and Szary, Andrzej},
  year = 2016,
  month = dec,
  journal = {\apj},
  volume = {833},
  number = {1},
  pages = {29},
  publisher = {The American Astronomical Society},
  doi = {10.3847/1538-4357/833/1/29}
}

@article{cpg+25,
  title = {A sibling of AR Scorpii: SDSS J230641.47+244055.8 and the observational blueprint of white dwarf pulsars},
  shorttitle = {A sibling of AR Scorpii},
  author = {Castro~Segura, N and Pelisoli, I and G{\"a}nsicke, B T and Coppejans, D L and Steeghs, D and Aungwerojwit, A and Inight, K and Romero, A and Sahu, A and Dhillon, V S and Munday, J and Parsons, S G and Kennedy, M R and Green, M J and Brown, A J and Dyer, M J and Pike, E and Garbutt, J A and Jarvis, D and Kerry, P and Littlefair, S P and McCormac, J and Sahman, D I and Buckley, D A H},
  year = 2025,
  month = oct,
  journal = {\mnras},
  volume = {543},
  number = {3},
  pages = {2116--2129},
  doi = {10.1093/mnras/staf1511}
}

@article{clm+05,
	title = {Arecibo timing and single-pulse observations of 17 pulsars},
	volume = {363},
	issn = {0035-8711},
	url = {https://doi.org/10.1111/j.1365-2966.2005.09499.x},
	doi = {10.1111/j.1365-2966.2005.09499.x},
	number = {3},
	urldate = {2023-07-04},
	journal = {\mnras},
	author = {Champion, D. J. and Lorimer, D. R. and McLaughlin, M. A. and Xilouris, K. M. and Arzoumanian, Z. and Freire, P. C. C. and Lommen, A. N. and Cordes, J. M. and Camilo, F.},
	month = nov,
	year = {2005},
	pages = {929--936},
}

@article{wws07,
	title = {The main-interpulse interaction of {PSR} {B1702}-19},
	volume = {467},
	issn = {0004-6361},
	url = {https://ui.adsabs.harvard.edu/abs/2007A&A...467.1163W},
	doi = {10.1051/0004-6361:20066957},
	urldate = {2024-08-20},
	journal = {\aap},
	author = {Weltevrede, P. and Wright, G. A. E. and Stappers, B. W.},
	month = jun,
	year = {2007},
	note = {ADS Bibcode: 2007A\&A...467.1163W},
	pages = {1163--1174},
}

@article{kwo+26,
  title = {The Thousand-Pulsar-Array programme on MeerKAT XIX: single-pulse data analysis, nulling and pulse energy distributions},
  shorttitle = {The Thousand-Pulsar-Array programme on MeerKAT XIX},
  author = {Keith, Michael J and Weltevrede, Patrick and Oswald, Lucy and Karastergiou, Aris and Song, Xiaoxi and Wang, Haoyue and Hsu, Jui-An and Johnston, Simon and Wright, Geoff and Bailes, Matthew and Serylak, Maciej},
  year = 2026,
  month = jul,
  journal = {\mnras},
  volume = {550},
  number = {1},
  pages = {stag1108},
  doi = {10.1093/mnras/stag1108}
}



\appendix

\section{Derivation of beat features involving harmonics and aliasing effect}

\subsection{Harmonic phases of periodic subpulse modulation}
\label{apdx:harmonic_phase}

Periodic subpulse modulation as a function of pulse number can be represented by a shifted Dirac comb of period $P_3$ convolved with a function $g$ describing the finite width and shape of the drift bands.
The Fourier transform of the Dirac comb is another Dirac comb.
A shift of the comb in the time domain results in a Fourier phase gradient as a function of frequency. Therefore a harmonic phase relation is obeyed, such that the $n$-th peak of the Dirac comb (at frequency $n/P_3$) in the Fourier domain has a phase that is $n$ times that at the fundamental, modulo $2\pi$.

Convolution with $g$ multiplies the complex Fourier amplitude of each peak by $G(n/P_3)$, where $G$ is the Fourier transform of $g$. If $G(n/P_3)$ is real and positive, no additional phase is introduced and thereby the harmonic phase relation of the shifted Dirac comb is preserved.
For a non-negative, even function $g$ (i.e. a symmetric drift band), the Fourier transform $G$ is real and positive over a range of low frequencies.
This range widens as $g$ narrows.
The harmonic phase relation therefore holds for frequencies of interest lying within this range.

As an example, take $g$ to be a rectangular function of width $W$ centred at zero.
Then $G$ is a sinc function, whose first zero is at $\pm 1/W$.
A duty cycle $W/P_3<1/N$ ensures that the fundamental and all harmonics up to order $N$ lie within the region where $G>0$.
Since the harmonics considered in this paper extend only to third order, a rectangular profile with a duty cycle below one-third suffices.
However, a rectangular function is a rather unnatural way to describe the finite width and shape of drift bands. A more realistic function $g$ would be a Gaussian centred at zero, whose Fourier transform is also Gaussian, hence real and positive at all frequencies.
Therefore, no restriction on the width is required to preserve the harmonic phase relation. 

Even if $G(n/P_3)$ would be negative, the result would be a phase offset of $\pi$. Although the simple harmonic phase relation would not necessarily hold, more complex phase relations between different harmonics can still be expected in scatter plots such as that shown in panel (1b) of \fig{fig:scatter-plot-F1-F2}.

\subsection{Beat relations involving harmonics and aliasing}
\label{apdx:beat_harmonic_alias}
As noted in \sect{sec:beat}, there is an equivalence between the beat features arising from the modulation of F1 (and its harmonics) with F2, and those resulting from its modulation with F3. Here, we provide a detailed derivation of this relationship by first working with the intrinsic (unaliased) harmonic spectral features and their associated beat features. After determining the intrinsic $1/P_2$ and $1/P_3$ frequencies alongside the Fourier phase of the spectral feature of interest, we account for the aliasing effect to determine the expected observable frequencies and phases.

We consider the established relation $\rm F1=F2+F3$ (see \sect{sec:beat}) and the intrinsic second harmonic of F1, i.e., $\hat{\mathrm{F1}}{}^{(2)} = 2\mathrm{F1}$. Here, the feature label (e.g., F1) is used as a generic variable applicable to both the $1/P_2$ and $1/P_3$ frequencies and the Fourier phase of the specific spectral feature. The hat notation ($\hat{\phantom{x}}$) designates the intrinsic spectral feature, whereas features without the hat represent the observed ones after aliasing is taken into account. 
If a spectral feature $\hat{\mathrm{F}}$ has an intrinsic $1/P_3$ frequency just above the Nyquist frequency ($0.5$~cpp), it results in an observed first-order aliased feature F. Its $1/P_3$ frequency is wrapped into the principal Nyquist range ($[0, 0.5]$~cpp), accompanied by inverted signs for its $1/P_2$ frequency and Fourier phase:
\begin{equation}
    \left\{
    \begin{aligned}
        (1/P_3)_{\mathrm{F}} &= 1 - (1/P_3)_{\hat{\mathrm{F}}} \\
        (1/P_2)_{\mathrm{F}} &= -(1/P_2)_{\hat{\mathrm{F}}} \\
        \theta_{\mathrm{F}} &= -\theta_{\hat{\mathrm{F}}}
    \end{aligned}
    \right. .
    \label{eq:aliasing}
\end{equation}
For example, since the intrinsic feature $\hat{\mathrm{F1}}{}^{(2)}$ has a $1/P_3$ frequency of $\rm\approx 0.66~cpp>0.5~cpp$, the observed aliased feature $\mathrm{F1}^{(2)}$ appears at $1/P_3 \approx 0.34$~cpp with opposite signs for its $1/P_2$ frequency and Fourier phase, as shown in panels (1a) and (1b) of \fig{fig:scatter-plot-F1-F2}.

Based on the interpretation involving the modulation of F1 (and its harmonics) with F2, the $\Sigma$ beat feature between F1 and F2, labelled $\hat{\mathrm{F4}}$, is given by:
\begin{equation}
    \hat{\mathrm{F4}} = \mathrm{F1} + \mathrm{F2} = \mathrm{F1} + (\mathrm{F1} - \mathrm{F3}) = 2\mathrm{F1} - \mathrm{F3} = \hat{\mathrm{F1}}{}^{(2)} - \mathrm{F3}.
    \label{eq:F4hat}
\end{equation}
This indicates that, in the unaliased domain, $\hat{\mathrm{F4}}$ can be equivalently interpreted as the $\Delta$ beat feature between $\hat{\mathrm{F1}}{}^{(2)}$ and F3.

It is worth noting that the designation of a $\Sigma$ or $\Delta$ beat associated with an aliased feature can be ambiguous because both $\hat{\mathrm{F4}}$ and $\hat{\mathrm{F1}}{}^{(2)}$ are subject to first-order aliasing.
Substituting the relations of \equ{eq:aliasing} into \equ{eq:F4hat}, we can rewrite the equation in terms of the observed spectral features:
\begin{equation}
    \mathrm{F4} = \mathrm{F1}^{(2)} + \mathrm{F3}.
    \label{eq:F4}
\end{equation}
This rearranged expression suggests that the observed F4 can be interpreted as the $\Sigma$ beat feature between the observed $\mathrm{F1}^{(2)}$ and F3, as suggested in panel (2a) of \fig{fig:scatter-plot-F1-F3}.
\equ{eq:F4hat} implies that the $\Delta$ beat calculated from the intrinsic $\hat{\mathrm{F1}}{}^{(2)}$ modulated by F3 undergoes aliasing to become observed F4. On the other hand, \equ{eq:F4} implies that one first determines the observed (aliased) $\mathrm{F1}^{(2)}$ and treats it as an `independent' feature that modulates with F3, yielding an unaliased $\Sigma$ beat feature F4. Ultimately, these two interpretations are equivalent; the apparent discrepancy in designating it as a $\Sigma$ or $\Delta$ beat depends on whether the aliasing effect is accounted for before or after evaluating the beats.
In the main text and Figs.~\ref{fig:scatter-plot-F1-F2} and \ref{fig:scatter-plot-F1-F3}, we adopt the interpretation in \equ{eq:F4}, as it provides a more intuitive approach for determining the locations of beat features in the 2DFS.
Following a similar logic, it can be demonstrated that any beat feature between F1 (or its harmonics) and F2 can be equivalently expressed as a beat feature between F1 (or its harmonics) and F3. Examples including F4, F5, F6, and F7 are presented in \tab{tab:beat_interpretations}, which summarizes all equivalent interpretations for each beat feature.

\begin{table*}
\caption{Summary of the equivalent interpretations of beat features arising from the modulation of F1 (and its harmonics) with F2 and F3. The intrinsic interpretation evaluates the beats prior to accounting for the aliasing effect, whereas the observed interpretation treats the aliased harmonic as an independent feature. Features denoted with a hat (e.g., $\hat{\mathrm{F1}}{}^{(2)}$) represent intrinsic spectral features before frequency wrapping.}
    \centering
    \renewcommand{\arraystretch}{1.5}
    \begin{tabular}{lccccc}
        \toprule
         \multirow{2}{*}{Feature} & \multirow{2}{*}{$1/P_3$ (cpp)} & \multicolumn{2}{c}{{Modulation with F2}} & \multicolumn{2}{c}{{Modulation with F3}} \\
        \cmidrule(lr){3-4} \cmidrule(lr){5-6}
        & & {Intrinsic interpretation} & {Observed interpretation} & {Intrinsic interpretation} & {Observed interpretation} \\
        \midrule
        F4 & 0.48 & aliased $\Sigma$ (F1 \& F2) & aliased $\Sigma$ (F1 \& F2) & aliased $\Delta$ ($\hat{\mathrm{F1}}{}^{(2)}$ \& F3) & unaliased $\Sigma$ ($\rm F1^{(2)}$ \& F3) \\
        F5 & 0.47 & unaliased $\Delta$ ($\hat{\mathrm{F1}}{}^{(2)}$ \& F2) & aliased $\Sigma$ (${\mathrm{F1}}{}^{(2)}$ \& F2) & unaliased $\Sigma$ (F1 \& F3) & unaliased $\Sigma$ (F1 \& F3) \\
        F6 & 0.15 & aliased $\Sigma$ ($\hat{\mathrm{F1}}{}^{(2)}$ \& F2) & unaliased $\Delta$ (${\mathrm{F1}}{}^{(2)}$ \& F2) & aliased $\Delta$ ($\hat{\mathrm{F1}}{}^{(3)}$ \& F3) & unaliased $\Sigma$ (${\mathrm{F1}}{}^{(3)}$ \& F3) \\
        F7 & 0.20 & aliased $\Delta$ ($\hat{\mathrm{F1}}{}^{(3)}$ \& F2) & unaliased $\Sigma$ (${\mathrm{F1}}{}^{(3)}$ \& F2) & aliased $\Sigma$ ($\hat{\mathrm{F1}}{}^{(2)}$ \& F3) & unaliased $\Delta$ (${\mathrm{F1}}{}^{(2)}$ \& F3) \\
        \bottomrule
    \end{tabular}
    \label{tab:beat_interpretations}
\end{table*}

The above demonstrates that the beat relations predict the same observable features, regardless of whether the observed (aliased) $\mathrm{F1}^{(2)}$ or the intrinsic $\hat{\mathrm{F1}}{}^{(2)}$ is considered. Similarly, the same observable features are predicted regardless of whether or not F1, F2, and F3 themselves are aliased.
Similar to the case considered in \tab{tab:beat_interpretations}, whether or not F1, F2 and F3 are aliased only affects the designation of the associated $\Sigma$ or $\Delta$ beat features.
For example, suppose that F1 is a first-order alias of an intrinsic $\hat{\mathrm{F1}}$ with a $1/P_3$ frequency of $\approx(1-0.33)=0.67~\rm cpp$ (see \equ{eq:aliasing}). 
In that case, the $\Sigma$ beat feature arising from the modulation between $\hat{\mathrm{F1}}$ and F2 will have a $1/P_3$ frequency of $\approx0.67+0.19=0.86~\rm cpp$.
This $\Sigma$ beat feature is therefore aliased to $1/P_3\approx0.14~\rm cpp$, coinciding exactly with F3. In contrast, when considering F1 to be unaliased as in  \sect{sec:beat}, the same observed F3 is designated as a $\Delta$ beat.
Similarly, whether or not F2 or F3 are aliased will result in the same spectral features.

\section{Additional analysis details}

\subsection{Summary of observations}
\label{apdx:obs_NF}

\tab{tab:obs} provides a summary of the observations used in our analysis, including the inferred nulling fraction (NF) as described in \sect{sec:null}.

\begin{table}
\caption{Observation parameters and the inferred nulling fraction (NF).
\label{tab:obs}}
\centering
\begin{tabular}{cccc}
\toprule
Date & No. rotations & No. antennas & NF (\%) \\
\midrule
2019-10-09 & 113 & 60 & 3.8 $\pm$ 2.0 \\
2019-12-16 & 1034 & 59 & 3.2 $\pm$ 0.6 \\
2020-02-24 & 201 & 28 & 3.0 $\pm$ 1.3 \\
2020-03-29 & 201 & 30 & 4.8 $\pm$ 1.8 \\
2020-05-13 & 202 & 29 & 0.5 $\pm$ 0.5 \\
2020-06-04 & 201 & 29 & 3.1 $\pm$ 1.3 \\
2020-07-03 & 201 & 29 & 9.9 $\pm$ 2.8 \\
2020-08-29 & 201 & 29 & 27.0 $\pm$ 3.1 \\
2020-09-25 & 200 & 29 & 0.5 $\pm$ 0.5 \\
2020-10-24 & 200 & 30 & 20.6 $\pm$ 2.8 \\
2020-11-20 & 201 & 30 & 11.6 $\pm$ 2.5 \\
2020-12-19 & 201 & 31 & 17.1 $\pm$ 3.0 \\
2021-01-18 & 201 & 29 & 2.5 $\pm$ 1.3 \\
2021-02-12 & 201 & 27 & 2.4 $\pm$ 1.5 \\
2021-03-11 & 201 & 28 & 22.2 $\pm$ 3.4 \\
2022-01-16 & 66 & 29 & 1.5 $\pm$ 1.5 \\
2022-02-19 & 67 & 31 & 37.9 $\pm$ 6.7 \\
2022-03-21 & 67 & 28 & 8.4 $\pm$ 4.9 \\
2022-04-17 & 67 & 32 & 1.5 $\pm$ 1.4 \\
2022-05-19 & 66 & 30 & 2.2 $\pm$ 2.0 \\
2022-06-14 & 66 & 31 & 1.5 $\pm$ 1.5 \\
2022-07-14 & 67 & 31 & 42.5 $\pm$ 20.9 \\
2022-08-11 & 66 & 29 & 9.7 $\pm$ 5.9 \\
2022-09-11 & 66 & 30 & 36.6 $\pm$ 6.5 \\
2022-10-10 & 67 & 29 & 4.1 $\pm$ 2.9 \\
2022-11-14 & 66 & 32 & 5.2 $\pm$ 2.8 \\
2022-12-16 & 66 & 31 & 3.0 $\pm$ 2.1 \\
2023-01-22 & 66 & 31 & 1.4 $\pm$ 1.4 \\
2023-02-18 & 66 & 31 & 4.2 $\pm$ 2.6 \\
2023-03-17 & 67 & 31 & 1.5 $\pm$ 1.4 \\
2023-05-15 & 66 & 32 & 1.5 $\pm$ 1.5 \\
2023-06-13 & 66 & 30 & 5.2 $\pm$ 2.9 \\
2023-07-12 & 66 & 30 & 1.4 $\pm$ 1.4 \\
2023-08-12 & 66 & 32 & 5.6 $\pm$ 2.8 \\
2023-09-13 & 65 & 30 & 3.2 $\pm$ 2.4 \\
2023-10-10 & 66 & 28 & 1.5 $\pm$ 1.5 \\
2023-11-13 & 66 & 29 & 11.6 $\pm$ 6.1 \\
2023-12-14 & 66 & 33 & 33.1 $\pm$ 24.5 \\
2023-12-16 & 67 & 33 & 11.8 $\pm$ 5.7 \\
2024-01-14 & 66 & 31 & 3.8 $\pm$ 3.1 \\
2024-02-26 & 67 & 31 & 1.4 $\pm$ 1.4 \\
2024-03-22 & 67 & 31 & 32.7 $\pm$ 5.8 \\
2024-04-16 & 67 & 25 & 5.8 $\pm$ 3.2 \\
2024-04-17 & 66 & 33 & 1.5 $\pm$ 1.4 \\
2024-05-21 & 67 & 32 & 2.7 $\pm$ 2.2 \\
2024-06-18 & 66 & 31 & 1.4 $\pm$ 1.4 \\
2024-09-16 & 67 & 30 & 2.9 $\pm$ 2.1 \\
2024-10-10 & 66 & 30 & 12.4 $\pm$ 5.0 \\
2024-11-16 & 66 & 29 & 5.1 $\pm$ 2.9 \\
2024-12-18 & 67 & 29 & 6.5 $\pm$ 3.2 \\
2025-01-11 & 66 & 30 & 1.9 $\pm$ 1.8 \\
2025-02-10 & 66 & 29 & 2.0 $\pm$ 1.8 \\
2025-03-20 & 66 & 30 & 54.7 $\pm$ 24.6 \\
2025-04-24 & 66 & 28 & 1.4 $\pm$ 1.4 \\
2025-05-26 & 67 & 30 & 6.7 $\pm$ 3.1 \\
2025-06-23 & 66 & 30 & 2.8 $\pm$ 2.1 \\
2025-07-29 & 67 & 30 & 2.0 $\pm$ 1.9 \\
2025-08-31 & 66 & 30 & 4.4 $\pm$ 3.2 \\
2025-10-06 & 67 & 29 & 7.2 $\pm$ 3.0 \\
2025-11-07 & 67 & 30 & 2.8 $\pm$ 2.3 \\
2025-12-06 & 67 & 31 & 3.0 $\pm$ 2.2 \\
2026-01-09 & 66 & 28 & 1.5 $\pm$ 1.4 \\
2026-02-14 & 66 & 30 & 17.5 $\pm$ 6.1 \\
2026-03-15 & 66 & 31 & 1.8 $\pm$ 1.9 \\
2026-04-26 & 66 & 28 & 5.4 $\pm$ 3.0 \\
\bottomrule
\end{tabular}
\end{table}

\subsection{Fourier phase analysis taking F2 and F3 to be fundamental}
\label{apdx:scatter-plot-F2-F3}

\begin{figure*}
    \centering
    \includegraphics[width=\linewidth]{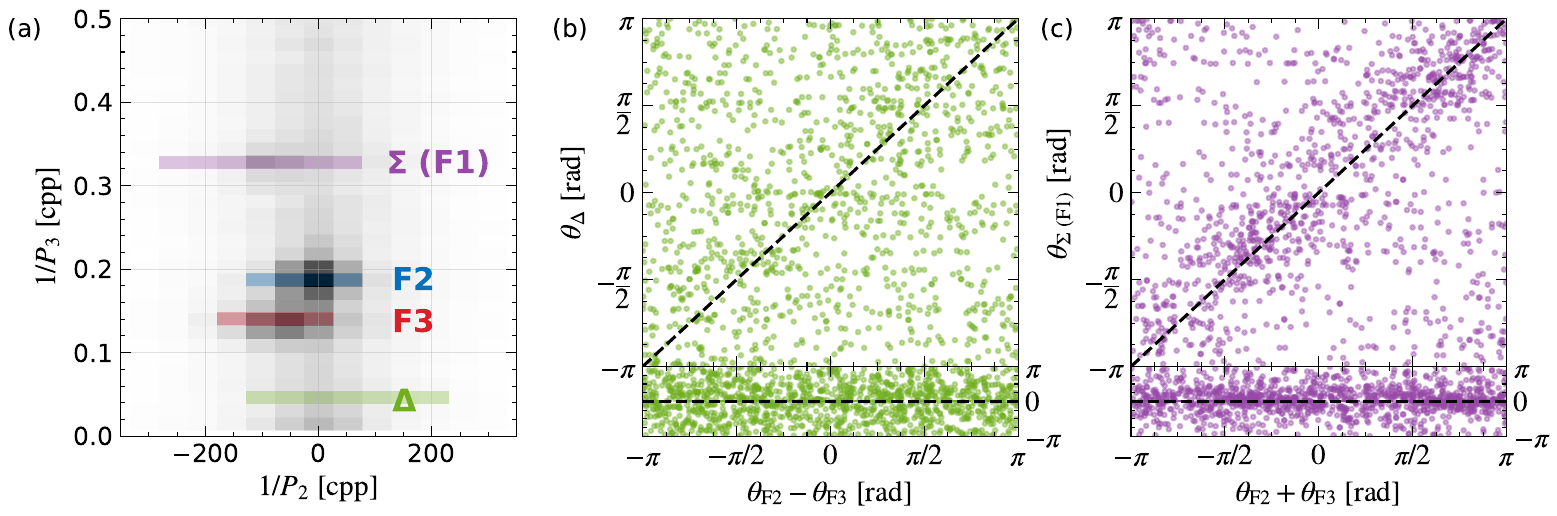}
    \caption{
    Similar to Figs.~\ref{fig:scatter-plot-F1-F2} and \ref{fig:scatter-plot-F1-F3}, but assuming that F2 and F3 modulate each other in C1, with $\Sigma$ beat feature as F1.
    This corresponds to (1f) in \fig{fig:corr-bar}.
    With 16 possible pixel combinations (four from F2 and four from F3), each scatter plot displays a total of 1232 individual data points.
    The circular correlation coefficients are $r_\Delta = 0.03$ (b) and $r_{\Sigma~\rm(F1)} = 0.24$ (c).
    }
    \label{fig:scatter-plot-F2-F3}
\end{figure*}

For completeness, similar to Figs.~\ref{fig:scatter-plot-F1-F2} and \ref{fig:scatter-plot-F1-F3}, \fig{fig:scatter-plot-F2-F3} presents the Fourier phase analysis under the assumption that F2 and F3 (both dominant features in C1) modulate each other. 

\subsection{Significance of Fourier phase correlations}
\label{apdx:additional_analysis}

\begin{figure*}
    \centering
    \includegraphics[width=\linewidth]{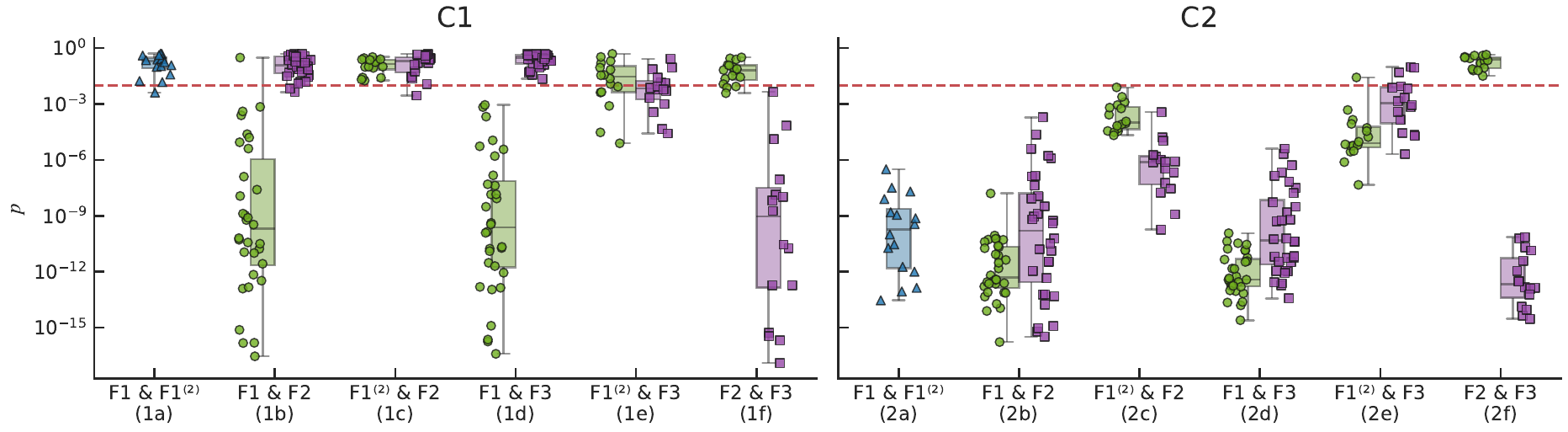}
    \caption{
    Strip plots displaying the distributions of $p$-values (one-sided test) for the detection of different spectral features. 
    The left-hand and right-hand panels show results in C1 and C2, respectively. Each data point represents a combination of pixels sampled from the shaded areas as defined in the 2DFS in Figs.~\ref{fig:scatter-plot-F1-F2}, \ref{fig:scatter-plot-F1-F3} and \ref{fig:scatter-plot-F2-F3}.
    For (1a) and (2a), the $p$-values represent the significance of correlation between F1 and its second harmonic $\rm F1^{(2)}$, indicated by blue triangles.
    All other groupings of spectral features show the significance of the $\Delta$ (green circles) and $\Sigma$ (purple squares) beat features between: (1b, 2b) F1 and F2; (1c, 2c) $\rm F1^{(2)}$ and F2; (1d, 2d) F1 and F3; (1e, 2e) $\rm F1^{(2)}$ and F3; and (1f, 2f) F2 and F3.
    Underlying box and whisker plots indicate the median, interquartile range, and the data range.
    For reference, the horizontal dashed red lines indicate the significance level of 0.01.
    }
    \label{fig:corr-bar}
\end{figure*}

We further quantify the observed correlation as shown in Figs.~\ref{fig:scatter-plot-F1-F2}, \ref{fig:scatter-plot-F1-F3}, and \ref{fig:scatter-plot-F2-F3} by calculating the corresponding $p$-values \citepalias[one-sided test, see][]{hww+25}.
For each pixel combination, for example, a pixel chosen from F1 and another from F2, a single $p$-value is calculated to assess the probability of the observed correlation in the Fourier phase scatter plot occurring purely by chance.
Since each spectral feature extends across multiple pixels, considering all possible pixel combinations yields a distribution of $p$-values, as presented in \fig{fig:corr-bar}.
The pixel combinations used here are drawn from Figs.~\ref{fig:scatter-plot-F1-F2}, \ref{fig:scatter-plot-F1-F3}, and \ref{fig:scatter-plot-F2-F3}.

For example, (2b) in \fig{fig:corr-bar} corresponds to panels (2a--2c) of \fig{fig:scatter-plot-F1-F2} which demonstrates consistency with F1 and F2 modulating each other in C2.
Both the purple squares ($\Sigma$), and especially the green circles ($\Delta$) are well below the horizontal dashed line (significance level of 0.01), providing compelling evidence that both beat features are detected significantly.
For all considered groupings of spectral features, the detection of one beat feature is always guaranteed, given the established $\rm F1=F2+F3$ relation (\sect{sec:subpulse_phase}). 
Since one always expects a second beat feature, its detection is crucial for verification of a given explored beat interpretation.
For example, the interpretation of F2 and F3 modulating each other in C2 is represented by (2f), which only shows a significant detection of the $\Sigma$ beat feature (purple squares). Therefore, this interpretation is disfavoured.

The left-hand panel of \fig{fig:corr-bar} summarizes the Fourier phase correlation results for C1. 
F2 and F3 are strong and of comparable strength in C1. Yet, as shown by (1f), there is no evidence that F2 and F3 modulate each other.
This can also be seen in the corresponding Fourier phase scatter plots, as shown in \fig{fig:scatter-plot-F2-F3}.

\begin{figure*}
    \centering
    \includegraphics[width=0.331\linewidth]{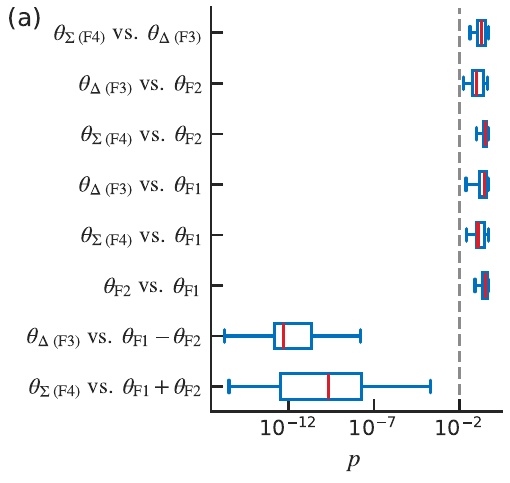}
    \includegraphics[width=0.331\linewidth]{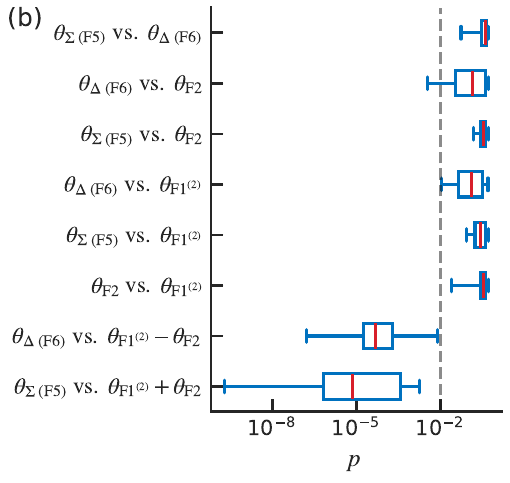}
    \includegraphics[width=0.331\linewidth]{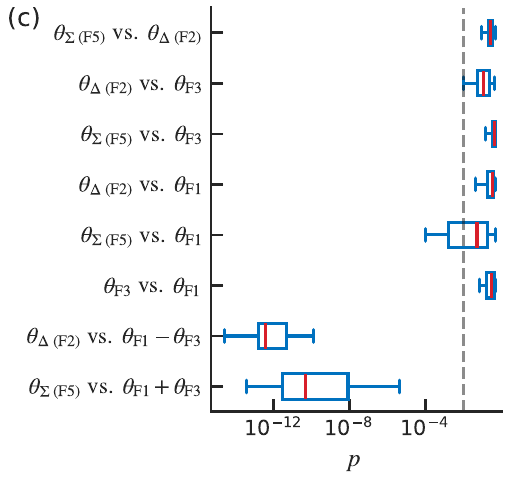} \\
    \includegraphics[width=0.331\linewidth]{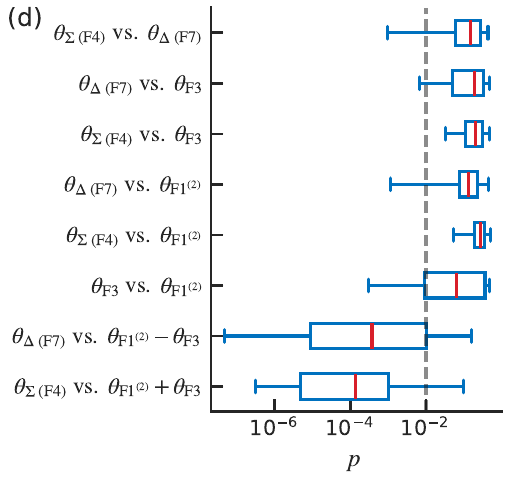}
    \includegraphics[width=0.331\linewidth]{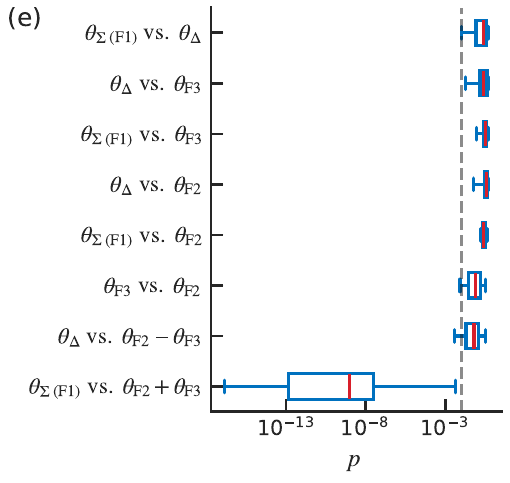}
    \caption{Box and whisker plots showing the distribution of the $p$-values (one-sided test) of the circular correlation coefficients, obtained by selecting different pixel combinations from the shaded regions in Figs.~\ref{fig:scatter-plot-F1-F2}, \ref{fig:scatter-plot-F1-F3} and \ref{fig:scatter-plot-F2-F3}.
    Panels (a) and (b) correspond to the scenarios presented in panels (2a) and (3a) of \fig{fig:scatter-plot-F1-F2}, respectively. 
    Panels (c) and (d) correspond to the scenarios presented in panels (1a) and (2a) of \fig{fig:scatter-plot-F1-F3}, respectively. 
    Panel (e) corresponds to the scenarios presented in panel (a) of \fig{fig:scatter-plot-F2-F3}. 
    For reference, the vertical dashed lines indicate the significance level of 0.01. In each panel, the bottom two distributions replicate those in \fig{fig:corr-bar}. The other rows should be centred at larger $p$-values, as the corresponding spectral features are expected to be uncorrelated.
    }
    \label{fig:J1901-rp-distribution}
\end{figure*}

The significance of the beat feature detections can be further assessed by comparing their $p$-value distributions as presented in \fig{fig:corr-bar} with those of the spectral features expected to be uncorrelated.
For example, the $p$-value distribution in panel (2b) of \fig{fig:corr-bar} is lower overall than those obtained for the correlations between any two spectral features expected to be uncorrelated, such as F1 and F2. This comparison is summarized in panel (a) of \fig{fig:J1901-rp-distribution}, which shows only small $p$-values in the bottom two rows.

The comparisons in \fig{fig:J1901-rp-distribution} are particularly useful when the $p$-values are not sufficiently small to provide strong evidence on their own. For example, although the $p$-values for the correlation shown in panel (2e) of \fig{fig:corr-bar}, corresponding to the correlations in panels (2b) and (2c) of \fig{fig:scatter-plot-F1-F3}, are relatively large, they are still much smaller than those obtained for spectral features expected to be uncorrelated as shown in panel (d) of \fig{fig:J1901-rp-distribution}. This provides evidence for a weak detection of these beat features.

\section{Feedback model}
\label{apdx:feedback}
\subsection{$k$ in \equ{eq:feedback}}
\label{apdx:k}

\begin{figure}
    \centering
    \includegraphics[width=\linewidth]{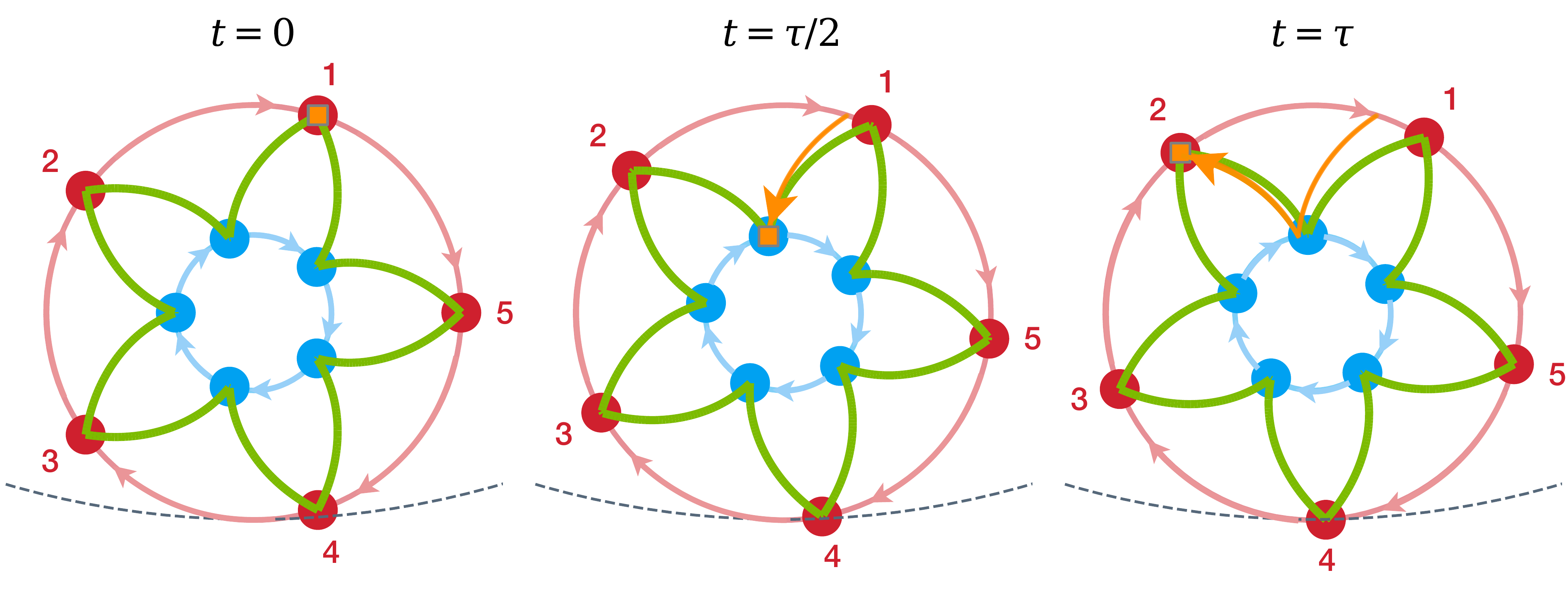}
    \caption{Similar to panel (b) of \fig{fig:feedback}, but here $k=1$, $n=5$, $P_5=25$ (corresponding to anticlockwise as indicated by the orange arrows), $\tau=4$ and constant interaction $m=0$ (uniform particle flow). According to \equ{eq:feedback}, the $P_3$ corresponding to the footprint circulation is $-20$ (clockwise as indicated by the red and blue arrows). 
    The footprints in the outer carousel (red) are numbered from 1 to 5.
    Three panels from left to right show the same configuration at three different times $t=0$, $t=\tau/2$, and $t=\tau$. One specific particle in the flow is highlighted by an orange square, and its trajectory is indicated by the orange curves.
    }
    \label{fig:time}
\end{figure}

\fig{fig:time} provides an illustrative example of the feedback model. The same configuration is shown at three different times $t=0$, $t=\tau/2$, and $t=\tau$ corresponding to the panels from left to right. The parameters in \equ{eq:feedback} are chosen such that a positive $P_5$ corresponds to an anticlockwise direction of circulation.
The resulting negative $P_3$ in \equ{eq:feedback} indicates that the footprints (red and blue) circulate in the opposite direction (clockwise), as indicated by the red and blue arrows.
If we track one specific particle (orange square) in the flow, it leaves the outer carousel (red) from footprint 1 at the time $t=0$ (left-hand panel) in the anticlockwise direction.
After a time interval of $\tau/2$, it arrives at the inner carousel (blue) and reflects back towards the outer carousel (middle panel). The particle follows the trajectory indicated by the orange curve.
Another $\tau/2$ later (at $t=\tau$), it returns to the outer carousel (red) at the location of footprint $2$. This corresponds to $k=1$.

We note that the angular velocity of the circulation of an individual particle (orange square) is different from that of the global pattern of the flow of particles (green curve). 
The former, driven by the $\boldsymbol{E} \boldsymbol{\times} \boldsymbol{B}$ drift, has a period of $P_5$.
In contrast, the latter is equivalent to the footprint circulation, which is essentially determined by how the collective distribution of particles in the flow forms a stable modulation pattern.

\subsection{$m=0$ illustration (\citetalias{rs75}-like scenario)}
\label{apdx:m=0}

\begin{figure}
    \centering
    \includegraphics[width=\linewidth]{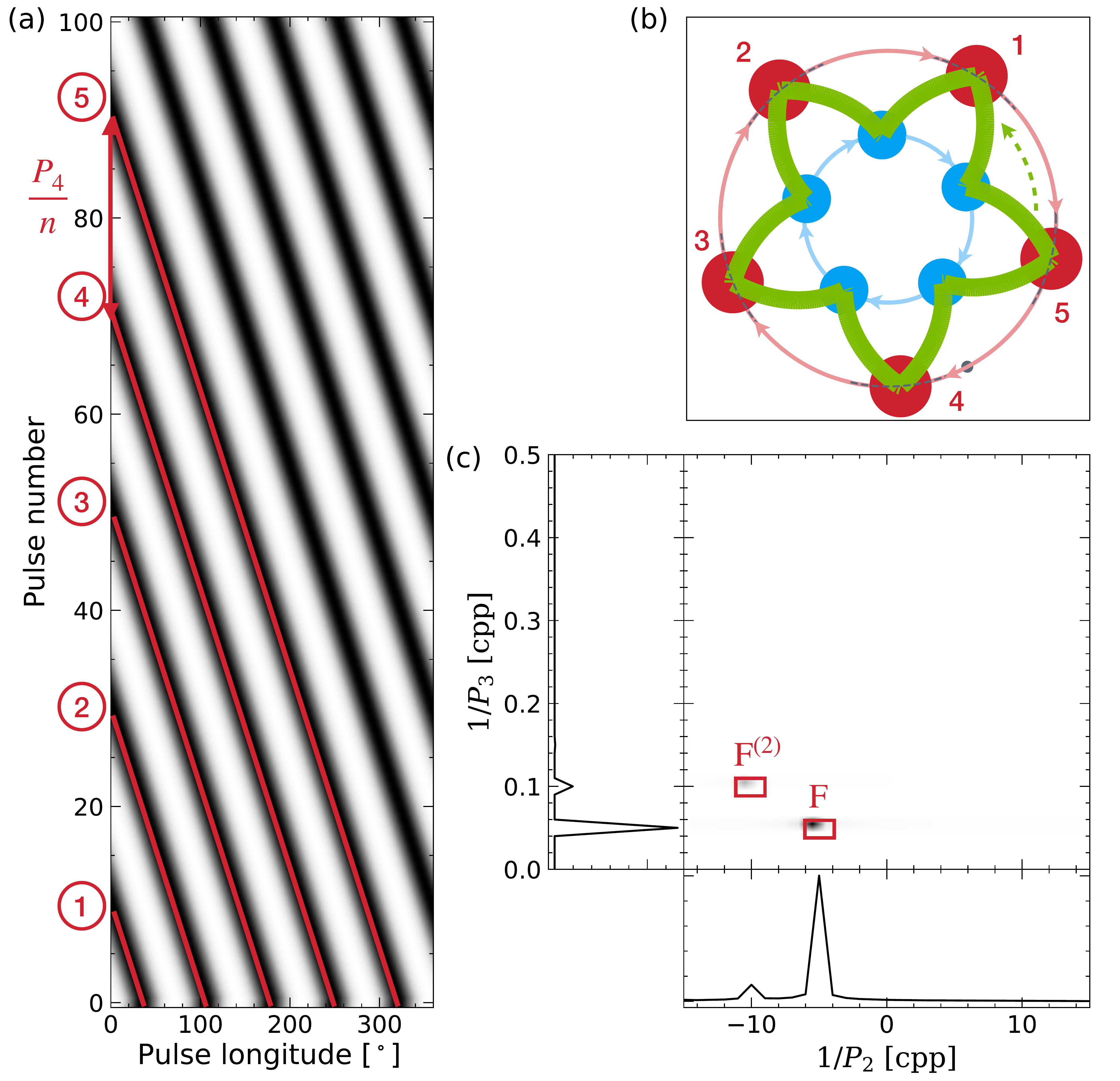} \\
    \includegraphics[width=\linewidth]{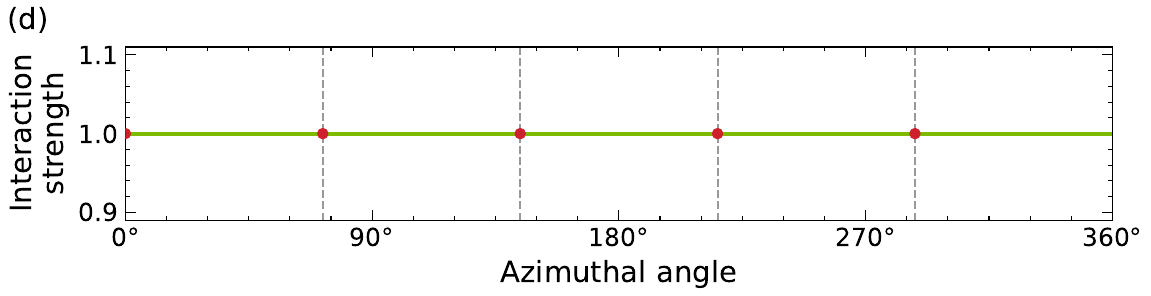}
    \caption{Similar to \fig{fig:feedback}, but here $k=1$, $n=5$, $P_5=25$ (anticlockwise as indicated by the dashed green arrows), $\tau=4$ and $m=0$. According to \equ{eq:feedback}, the $P_3$ corresponding to the footprint circulation is $-20$ (clockwise as indicated by the blue and red arrows). 
    In the pulse stack in panel (a), the drift band corresponding to a given footprint $i$ is indicated by the solid red line labelled $i$.
    In the 2DFS in panel (c), the fundamental (first harmonic) and its second harmonic features are labelled as F and $\rm F^{(2)}$, respectively.
        }
    \label{fig:m=0}
\end{figure}

An example of a feedback model with constant interaction $m=0$ (uniform particle flow) is illustrated in panel (b) of \fig{fig:m=0}. 
This setup is identical to that in \fig{fig:time}.
The simulation parametrises an aligned rotator, i.e. a magnetic axis aligned with the rotation axis. This implies that the line of sight always intersects the outer carousel (red).
The particles (green) are chosen to circulate in the anticlockwise direction, as indicated by the dashed green arrow, corresponding to a positive $P_5=25$.
According to \equ{eq:feedback}, given $n=5$, $k=1$ and $\tau=4$, the resulting $P_3$ associated with the footprint circulation is $P_3=-20$, where the negative sign denotes an opposite circulation in the clockwise direction, as indicated by the red and blue arrows.

Because of the constant interaction as shown in panel (d) of \fig{fig:m=0}, the emission produced by different footprints is identical and constant, resulting in a series of diagonal drift bands with equal intensity, as indicated by the solid red lines in panel (a).
These circulating footprints with constant intensity are reminiscent of $n$ sparks circulating with a period of $P_4=nP_3$ in the \citetalias{rs75} model, which only results in a single spectral feature F (and its harmonics, e.g. $\rm F^{(2)}$), as shown in the 2DFS in panel (c).

\subsection{Footprint modulation}
\label{apdx:footprint}

For $n$ footprints circulating with a period of $P_4$, the azimuthal angle of the $i$-th footprint at the $j$-th pulse number is given by:
\begin{equation}
    \phi_i(j) = \frac{2\pi}{n}i + \frac{2\pi}{P_4}j.
    \label{eq:phi}
\end{equation}
The particle flow circulates with a period of $P_5$, carrying a periodic modulation pattern that completes $m$ full cycles in a complete circle.
The phase of this modulation pattern as a function of azimuthal angle $\phi$ and time (pulse number) $j$ can be written as:
\begin{equation}
    \Phi(\phi, j) = m\phi - 2\pi\frac{m}{P_5}j.
    \label{eq:Phi}
\end{equation}
This phase determines the interaction strength experienced by the footprints.
Consequently, at any given azimuthal angle $\phi$ (or equivalently, a specific pulse longitude), the interaction strength is periodically modulated with a period $P_5/m$.

Alternatively, if we track an individual footprint, the modulation phase ($\Phi$) it experiences, and hence its intensity, varies as a function of time (parametrised by $j$).
To determine the modulation phase observed by a specific footprint, we substitute its time-dependent position $\phi_i(j)$ (\equ{eq:phi}) into the modulation phase $\Phi(\phi, j)$ (\equ{eq:Phi}):
\begin{align}
    \Phi_i(j)
    &= m \left( \frac{2\pi}{n}i + \frac{2\pi}{P_4}j \right) - 2\pi\frac{m}{P_5}j = 2\pi \frac{m}{n} i + 2\pi m \left( \frac{1}{P_4} - \frac{1}{P_5} \right)j \nonumber \\
    &= 2\pi \frac{m}{n} i - 2\pi \frac{mk}{n\tau} j.
\label{eq:modulation_phase}
\end{align}
The final step uses \equ{eq:feedback} and $P_4=nP_3$.
Consequently, a given footprint (given $i$) experiences an amplitude modulation with a period of $n\tau/(m|k|)$, as used in \sect{sec:feedback} in the main text.

\subsection{$m\bmod n \neq 0$ illustration}
\label{apdx:m=1}

\begin{figure}
    \centering
    \includegraphics[width=\linewidth]{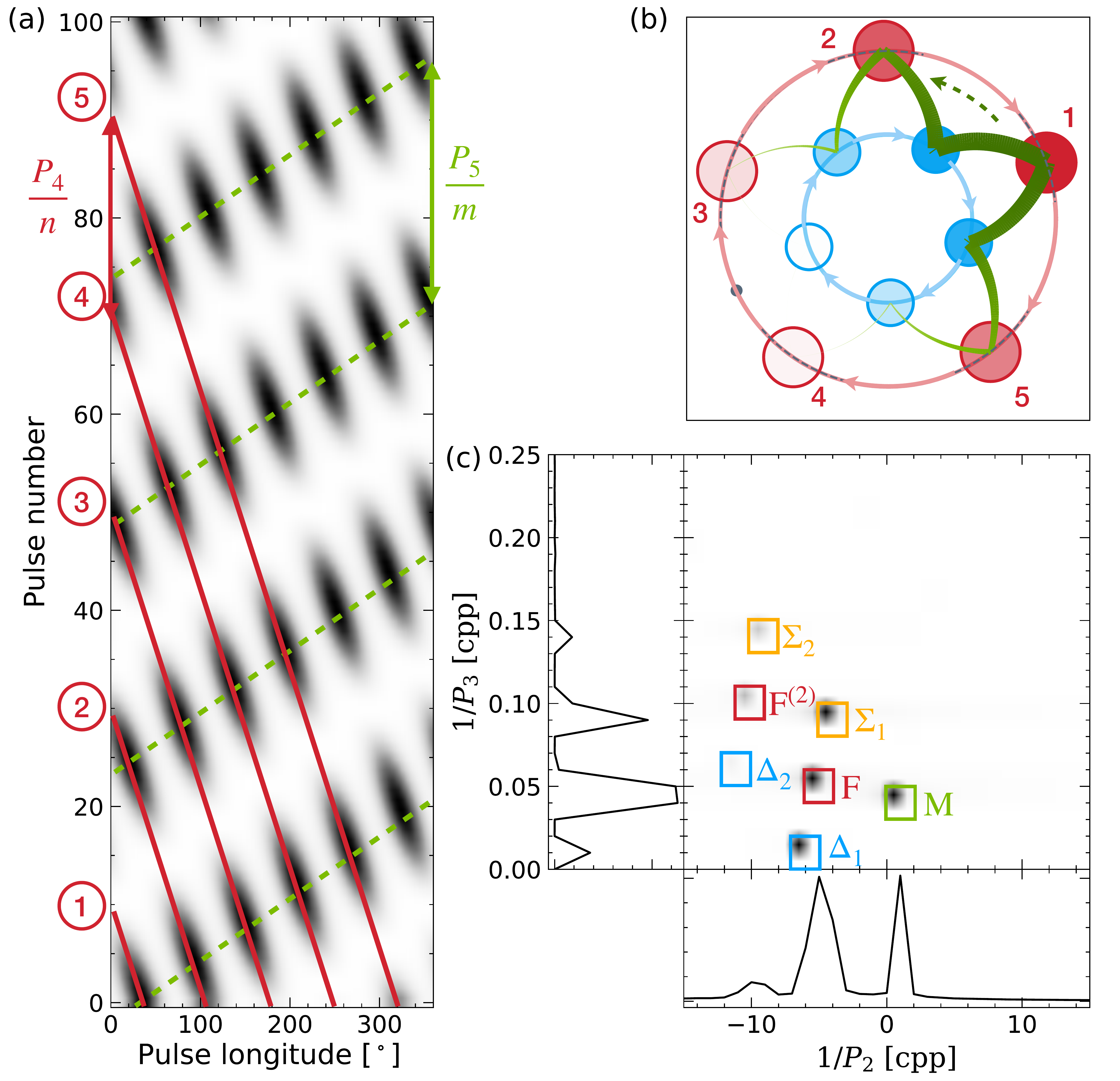}
    \includegraphics[width=\linewidth]{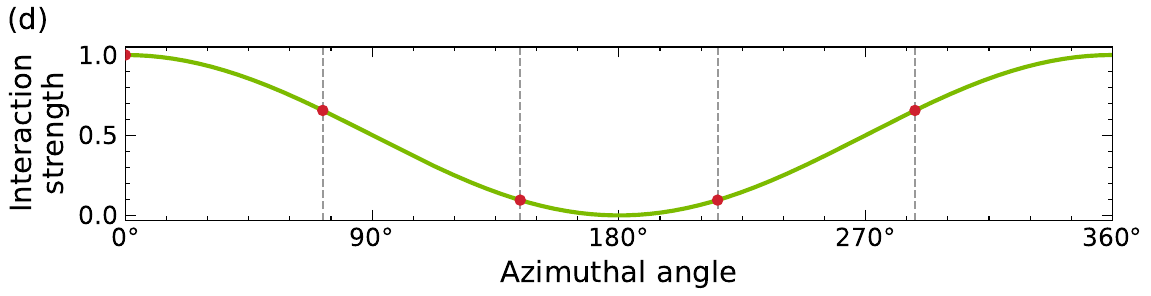}
    \caption{Similar to \fig{fig:m=0}, but here $m=1$.
    }
    \label{fig:m=1}
\end{figure}

Consider the case illustrated in \fig{fig:m=1} where $n=5$ and $m=1$.
Because the modulation phase $\Phi_i(j)$ in \equ{eq:modulation_phase} depends on the footprint index $i$, different footprints experience this modulation at different phases. 
The points of constant modulation phase form a series of diagonal stripes across the pulse stack, represented by the dashed green lines in panel (a), with a $P_3=P_5/m$ and a $|P_2|$ about $n/m$ times larger than that of the footprint drift bands (solid red lines)\footnote{$|P_2|$ will deviate from this when circulation during time $|P_2|$ is significant. However, in the simulations here and in \fig{fig:feedback} this effect is small enough to not affect any of the conclusions drawn from this ratio.}.
This corresponds to the modulation feature (labelled as M) in panel (c) of \fig{fig:m=1}.
In this aligned rotator simulation, the entire outer carousel (red) is visible.
Consequently, the five subpulses originating from the five footprints are evenly distributed in pulse longitude, resulting in a footprint circulation with $|P_2|\simeq360^\circ/5=72^\circ$.
Since $m=1$, a complete sweep of the line of sight observes only one bright and one weak modulation phase per rotation, corresponding to $|P_2|=360^\circ$.
In the absence of the aliasing effect, the sign of this $P_2$, which quantifies the observed drift direction, is the same as that of $P_5$ in \equ{eq:feedback}.

Mathematically speaking, the resulting pulse stack is the unmodulated pulse stack with continuous drift bands (solid red lines) multiplied by a diagonal modulation pattern (dashed green lines) characterized by the $P_2$ and $P_3$ associated with the modulation particle flow.
According to the convolution theorem, this multiplication in the time domain corresponds to a convolution in the Fourier domain. 
Consequently, the 2DFS in panel (c) of \fig{fig:m=1} exhibits a series of beat features ($\Sigma$ and $\Delta$), which emerge from the convolution of the footprint drift features (F and its harmonics) with the modulation feature (M).
The subscripts of $\Sigma$ and $\Delta$ denote the harmonic order of the footprint drift feature involved.

\subsection{$m\bmod n = 0$ illustration}
\label{apdx:m=n}

\begin{figure}
    \centering
    \includegraphics[width=\linewidth]{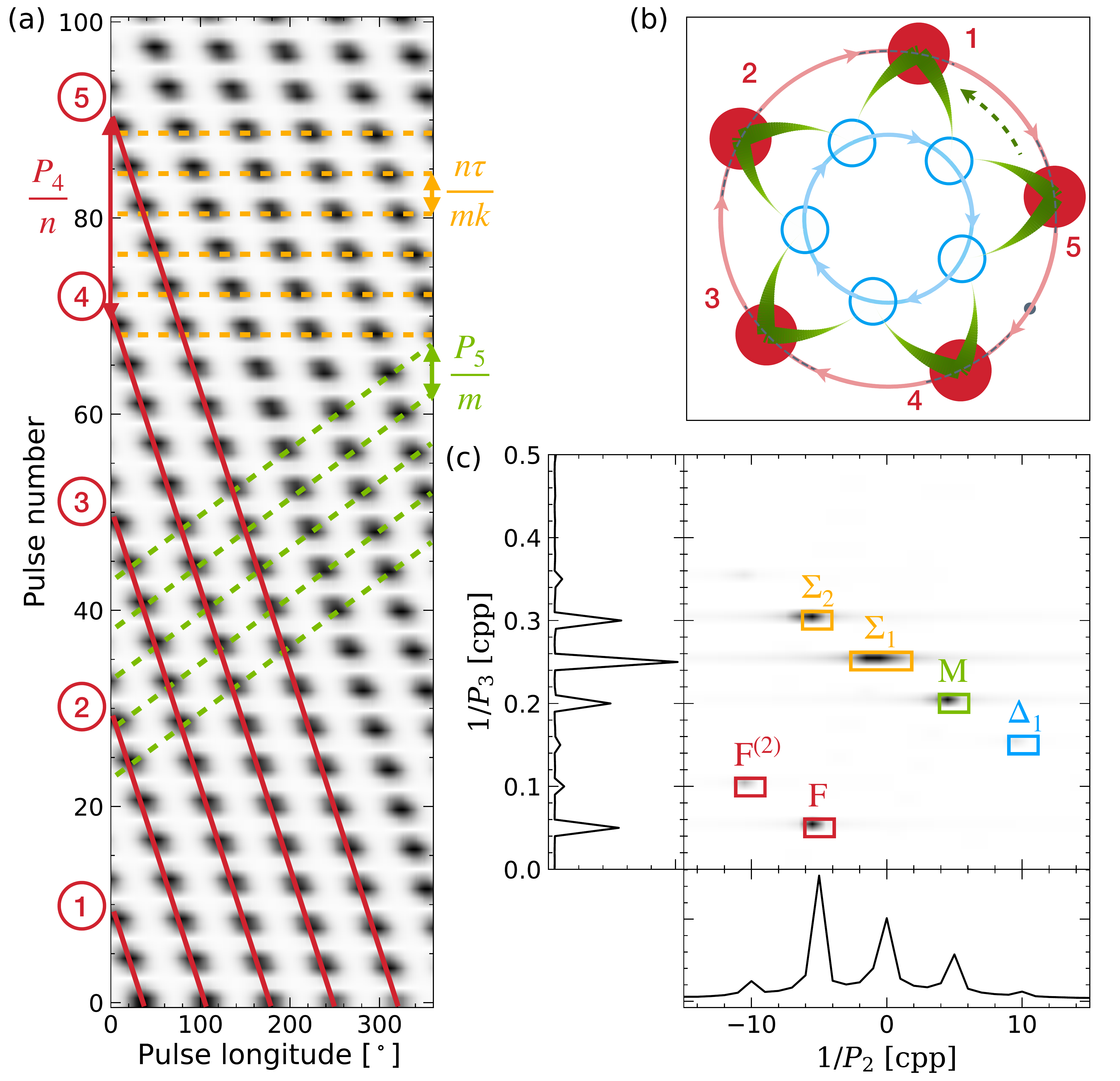}
    \includegraphics[width=\linewidth]{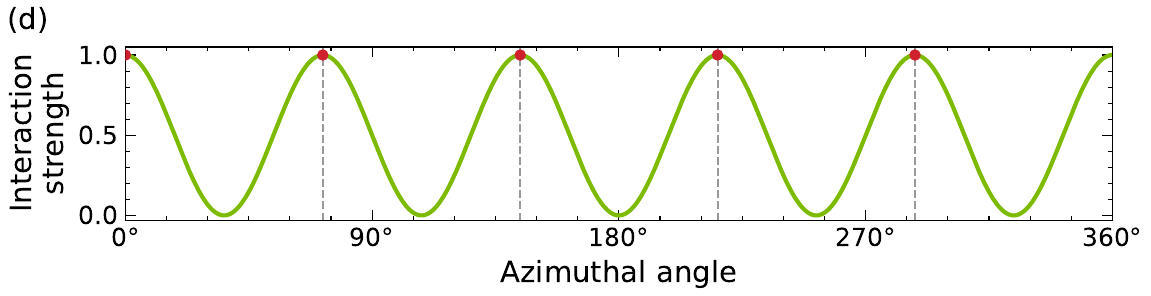}
    \caption{Similar to \fig{fig:m=1}, but here $m=5$.}
    \label{fig:m=5}
\end{figure}

By increasing $m$ from $m=1$ (\fig{fig:m=1}) to $m=n$, the $1/P_2$ and $1/P_3$ frequencies of the modulation feature (M) are scaled by a factor of $n$, as shown in panel (c) of \fig{fig:m=5}.
Under this specific condition, the $\Sigma$ beat feature between this modulation feature (M) and the footprint drift feature (F) corresponds to an amplitude modulation feature $\Sigma_1$ with a period of $\tau$, the time it takes for particles to stream back and forth between the two carousels.
In the following, we demonstrate that for cases where $m$ is an integer multiple of $n$, equivalent results are obtained when interpreting F (and its harmonics) as being modulated by $\Sigma_1$, as when F is modulated by M.

When $m$ is an integer multiple of $n$, the first term in \equ{eq:modulation_phase} simplifies to an integer multiple of $2\pi$. Since adding an integer multiple of $2\pi$ has no physical effect on the modulation phase, this term can be discarded. 
As a result, the modulation phase does not depend on the footprint index $i$; i.e., all footprints experience the same modulation phase at any given time and are therefore synchronized.

When F is interpreted as being modulated by M, the constant modulation phases correspond to the diagonal dashed green lines in panel (a) of \fig{fig:m=5} with $P_3=P_5/m$. 
However, when this modulation pattern is sampled by the footprint drift bands (solid red lines), the resulting emission pattern becomes mathematically equivalent to a pure amplitude modulation with a period of $\tau$ (horizontal dashed yellow lines). 
The $P_3$ of this amplitude modulation, i.e. the vertical spacing of the horizontal dashed yellow lines, is $n\tau/(m|k|)$, which is the beat period between the footprint $P_3$ and the modulation period $P_5/m$ according to \equ{eq:feedback} given $m=n$ and $k=1$. This modulation feature corresponds to $\Sigma_1$.
Because the footprint drift feature (F) inherently contains a series of harmonics, both modulation perspectives yield identical features in the 2DFS.
For instance, in panel (c), the $\Sigma_2$ beat feature between the second harmonic of the footprint drift ($\rm F^{(2)}$) and the modulation feature (M) can be equivalently identified as the $\Sigma$ feature between this amplitude modulation feature $\Sigma_1$ and the footprint drift feature (F).
Therefore, this modulation pattern can be equivalently described by F being modulated by M, as well as by $\Sigma_1$.

\subsection{Non-sinusoidal modulation}
\label{apdx:non-sinusoidal}
Although the modulation of the particle flow is assumed to be purely sinusoidal as a function of azimuthal angle in the simulations presented here, as shown in panels (d) of Figs.~\ref{fig:m=1} and \ref{fig:m=5}, a more general case can be considered.
Fundamentally, any arbitrary modulation pattern can be mathematically decomposed into a linear combination of Fourier components, each with its own weight and relative phase offset.
However, since the particle flow represents a physical interaction strength, it cannot be negative.
Therefore, rather than viewing the modulation pattern merely as a sum of abstract mathematical sinusoidal terms, it is physically more intuitive to describe it as a superposition of independent, non-negative sinusoidal modes.
Each mode corresponds to a specific integer $m$ and consists of a trigonometric Fourier component combined with a sufficient zero-frequency (constant) component to ensure its non-negativity.
Because this formulation is mathematically equivalent to simply redistributing the overall zero-frequency component of a standard Fourier series, such non-negative sinusoidal modes are equally complete and can thus represent any arbitrary physical modulation pattern.
Under the assumption of a linear relationship between the particle flow interaction and the resulting intensity produced by the footprints, the total observed pulse stack is simply the sum of the contributions from each non-negative sinusoidal mode.
By virtue of the linearity of the Fourier transform, the final 2DFS of the total pulse stack is equivalent to the coherent sum of the complex 2DFS associated with each individual sinusoidal mode.
Consequently, although more spectral features may emerge in the case of non-sinusoidal modulation, the $1/P_2$ and $1/P_3$ frequencies for a given $m$-sinusoidal mode adhere precisely to the relations described above.


\bsp	
\label{lastpage}
\end{document}